\documentclass{elsart}
\pdfoutput=1
\usepackage{amsmath,amssymb,amsfonts,graphicx}
\usepackage[scaled=0.9]{helvet}
\usepackage{microtype}
\usepackage{url,color,subfigure}
\newcommand{\mnote}[1]{%
}
\newtheorem{definition}{Definition}

\newtheorem{lemma}[definition]{Lemma}

\newtheorem{observation}[definition]{Observation}

\newcommand{\re}{\operatorname{Re}}

\newcommand{\spec}{\operatorname{spec}}
\newcommand{\sgn}{\operatorname{sign}}

\newcommand{\ef}{\operatorname{e}}
\newcommand{\R}{\mathbb{R}} %{I\!\!R}

\renewcommand{\epsilon}{\varepsilon}
\renewcommand{\phi}{\varphi}

\begin{document}
\begin{frontmatter}
  \title{Small-scale instabilities in dynamical systems with sliding}
\author[po]{J.~Sieber\corauthref{cor}}
\corauth[cor]{Corresponding author.}
\ead{jan.sieber@port.ac.uk} and
\author[um]{P.~Kowalczyk}
\address[po]{Department of Mathematics,
  University of Portsmouth, Portsmouth, PO1 3HF, U.K.}
\address[um]{School of Mathematics,
  University of Manchester, Manchester, M13 9PL, U.K.}

\begin{abstract}
  We demonstrate with a minimal example that in \emph{Filippov}
  systems (dynamical systems governed by discontinuous but piecewise
  smooth vector fields) stable periodic motion with sliding is not
  robust with respect to stable singular perturbations. We consider a
  simple dynamical system that we assume to be a quasi-static
  approximation of a higher-dimensional system containing a fast
  stable subsystem. We tune a system parameter such that a stable
  periodic orbit of the simple system touches the discontinuity
  surface: this is the so-called grazing-sliding bifurcation. The
  periodic orbit remains stable, and its local return map becomes
  piecewise linear. However, when we take into account the fast
  dynamics the local return map of the periodic orbit changes
  qualitatively, giving rise to, for example, period-adding cascades
  or small-scale chaos.
\end{abstract}
\end{frontmatter}

\section{Introduction}
\label{sec:intro}
Filippov systems are dynamical systems governed by discontinuous but
piecewise smooth ordinary differential equations (ODEs) and they
occur, for example, in the modelling of mechanical systems (with dry
friction) \cite{VB99}, of electrical systems with switching
\cite{BV01}, or of population dynamics (when, depending on the
population size, either the individuals switch habitat or diet, or
harvesting becomes restricted \cite{DGR07}). The phase space of a
Filippov system is partitioned into domains where for each domain a
different ODE governs the dynamics. In the simplest case we have two
domains such that
\begin{equation}
  \label{eq:filip}
  \dot x=
  \begin{cases}
    f_-(x) & \mathrm{if\ } h(x)<0\mbox{,}\\
    f_+(x) & \mathrm{if\ } h(x)\geq0\mbox{.}
  \end{cases}
\end{equation}
The boundary between the domains is called the
\emph{switching manifold}: $\mathcal{ H}_s=\{h(x)=0\}$.

\begin{figure}[t]
  \centering
  \includegraphics[width=\textwidth]{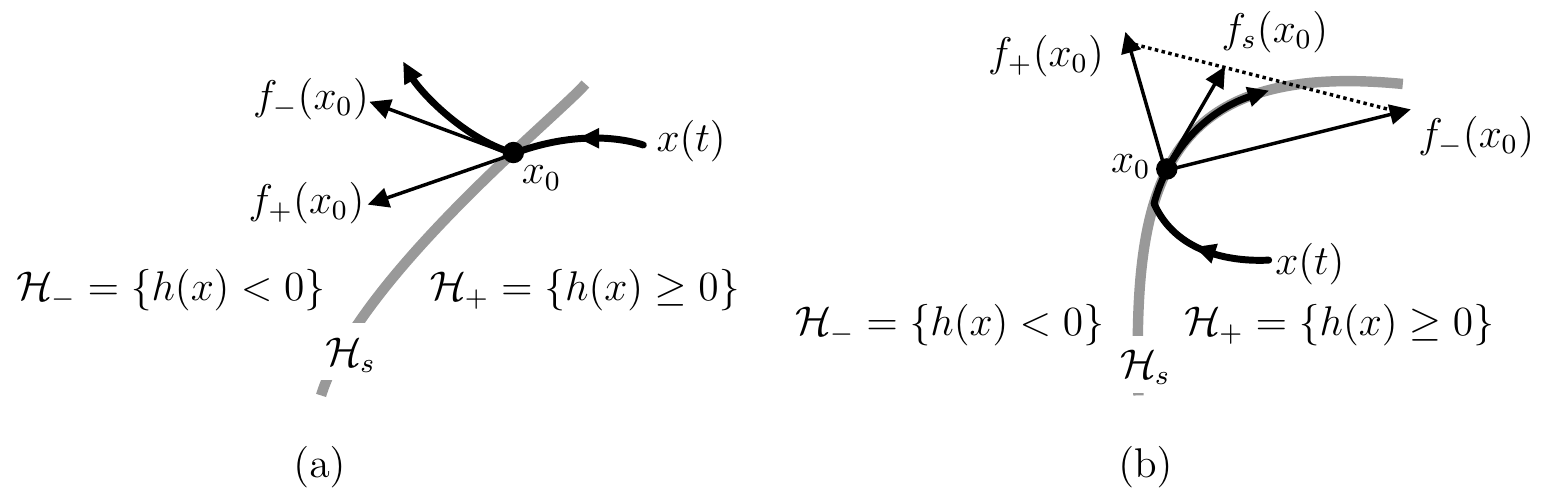}
  \caption{A trajectory $x(t)$ through a point $x_0$ on the switching
    manifold $\mathcal{H}_s$ generically either (a) crosses the
    switching manifold $\mathcal{H}_s$ if the vector fields $f_+$ and
    $f_-$ both point in the same direction relative to
    $\mathcal{H}_s$, or (b) slides along $\mathcal{H}_s$ if $f_+$ and
    $f_-$ both point toward $\mathcal{H}_s$.}
  \label{fig:slidesketch}
\end{figure}

A special feature of Filippov systems is the so-called \emph{sliding mode},
which means that a trajectory of the system does not follow any of the
ODEs governing the domains but it rather `slides' along the switching
manifold, following a convex combination of the ODEs governing the
adjacent domains:
\begin{equation}
  \label{eq:slide}
  \dot x=f_s(x)=\frac{[\partial h(x)f_-(x)]\cdot f_+(x)-
    [\partial h(x)f_+(x)]\cdot f_-(x)}{
    \partial h(x)[f_-(x)-f_+(x)]}\mbox{,}
\end{equation}
where `$\partial$' denotes partial differentiation (in the current case with respect to the variable $x$).
Sliding occurs in all points $x_0$ of the switching manifold
$\mathcal{H}_s$ where both vector fields point toward $\mathcal{H}_s$,
that is, $\partial h(x)f_+(x)<0$ and $\partial h(x)f_-(x)>0$.
Figure~\ref{fig:slidesketch} illustrates the two typically observed
cases of interaction between flows and switching manifold.

An important question for modelling is how sliding in a Filippov
system is affected by perturbations. If we add a small perturbation to
$f_-$ or $f_+$ or to the switching decision $h$ (the derivative of the
perturbation is also assumed to be small) then any exponentially
stable periodic orbit or equilibrium of a Filippov system persists
\cite{BBCK07} and remains stable. This also applies to
pseudo-equilibria (equilibria of the sliding flow $f_s$, sitting
exactly on the switching manifold) and to periodic orbits that have
sliding segments. This persistence mirrors the results of classical
bifurcation and invariant manifold theory for smooth dynamical systems
\cite{F79}.

Another typical perturbation arising in the modelling process are
\emph{stable singular} perturbations.  In a simple model one has
replaced rapidly converging parts of the dynamics with their
equilibrium, making the assumption that this equilibrium follows the
slow dynamics \emph{quasi-statically}. In a more complex of the same
system (or in reality) the equilibrium of the fast dynamics is not
attained perfectly, which constitutes a small perturbation. Practical
examples of this type of perturbation are small capacitances and
inductances in electrical circuits, imperfect rigidity in mechanical
systems, or fast chemical reactions (or other processes) in biological
systems. Again, for smooth dynamical systems classical theory
\cite{F79} proves that all hypberbolic equilibria, periodic orbits
and, more generally, normally hyperbolic invariant manifolds persist.
That is, for example, an exponentially stable equilibrium or periodic
orbit (and any of its bifurcations) observed in a simple model
obtained by making quasi-static assumptions is also present when the
fast dynamics is taken into account as long as the difference in time
scale is sufficiently large. In general, in smooth dynamical systems
any phenomenon that persists under regular perturbations
(perturbations of the right-hand-side) also persists under stable
singular perturbations. Fenichel's Theorem reduces hyperbolic singular
perturbations to regular ones by proving the existence of a normally
hyperbolic invariant manifold \cite{F79}.

In order to find general statements how the dynamics of Filippov
systems is affected by stable singular perturbations one has to study
slow-fast systems of the form
\begin{align}
  \dot x&=\
  \begin{cases}
    f_-(x,y,\epsilon) & \mathrm{if\ } h(x,y,\epsilon)<0\mbox{,}\\
    f_+(x,y,\epsilon) & \mathrm{if\ } h(x,y,\epsilon)\geq0,
  \end{cases}\label{eq:sgfilipslow}\\
  \epsilon\dot y &=\ g(x,y,\epsilon)\label{eq:sgfilipfast}\mbox{.}
\end{align}
In system \eqref{eq:sgfilipslow},\,\eqref{eq:sgfilipfast} $\epsilon$ measures
the difference in the time scales between the evolution of the slow
variable $x$ and the evolution of the fast variable $y$. We assume
that for $\epsilon=0$ the (then algebraic) equation
\eqref{eq:sgfilipfast} can be solved for $y$ for all $x$, and that
this solution $y_0(x)$ satisfies the stability condition
\begin{align}
%  0&=\ g(x,y_0(x),0) \label{eq:y0solv}\\
  0&>\ -c>\re\spec\partial_2g(x,y_0(x),0)\label{eq:y0stab}
\end{align}
with a uniform decay rate $c$ (subscript $2$ in the differential operator indicates differentiation with respect to the second argument).
Condition~\eqref{eq:y0stab} means that
$y_0(x)$ is a locally exponentially stable equilibrium of the fast
subsystem \eqref{eq:sgfilipfast} if we treat the variable $x$ in
\eqref{eq:sgfilipfast} as a parameter  and set $\epsilon$ to $0$ in the
right-hand-side of \eqref{eq:sgfilipfast}.

Let us assume that the quasi-static approximation (called the
\emph{reduced} system from now)
\begin{equation}
  \label{eq:filipred}
  \dot x=
  \begin{cases}
    f_-(x,y_0(x),0) & \mathrm{if\ }
    h(x,y_0(x),0)<0\mbox{,}\\
    f_+(x,y_0(x),0) & \mathrm{if\ }
    h(x,y_0(x),0)\geq0
  \end{cases}
\end{equation}
has a stable periodic orbit $x(t)$ which is switching in $x_0=x(t_0)$
from the subdomain $\{x: h(x,y_0(x),0)\geq0\}$ to sliding inside the
manifold $\{x: h(x,y_0(x),0)=0\}$ (as shown in
Figure~\ref{fig:slidesketch}(b)). What happens to this periodic orbit
if we include the singular perturbation effects by changing $\epsilon$
to a positive value?

The references \cite{F02,F02a} studied this question under the
simplifying assumption that the switching decision does not depend on
the fast variable $y$ (that is, $\partial_2h=0$ in
\eqref{eq:sgfilipslow},\,\eqref{eq:sgfilipfast}) and proved that
stable periodic orbits persist when $\epsilon$ becomes positive but
they acquire small boundary layers after switching.
%The assumption
%$\partial_2h=0$ corresponds to the special case $\theta=1$ in example
%\eqref{eq:peqslow},\,\eqref{eq:peqfast}.
In practice, it is often difficult to check or guarantee the condition
$\partial_2h=0$.  For example, a switch in an electronic circuit may
depend on the voltage potential at a node $j$ which is only a function
of neighboring node potentials, which (let us assume) are all slow
variables.  However, taking into account a small parasitic capacitance
(of order $\epsilon$) affecting node $j$ will change its voltage
potential into a fast variable, thus, making the switching decision
dependent on a fast variable.  The same effect occurs more generally
in modelling: one often studies the simple model ($\epsilon=0$) only,
and it is impossible to tell if switching decisions depend on fast
variables for any of the possibly significant singular perturbations
without studying the more complex model ($\epsilon>0$). This means
that the assumption $\partial_2h=0$ of \cite{F02,F02a} is
mathematically convenient because it allows to prove persistence of
stable periodic orbits but it does not cover all cases of practical
interest.

Our paper studies the problem how the dynamics near a periodic orbit
changes under a stable singular perturbation if the switching decision
depends also on fast variables: $\partial_2 h\neq0$ in
\eqref{eq:sgfilipslow}. It turns out that there are qualitative
changes if we increase $\epsilon$ from $0$ to a positive
value. Possible scenarios are, for example, a period-adding cascade on
a parameter range of order $\epsilon$, or small-scale chaos (also of
order $\epsilon$) around a periodic orbit that is exponentially stable
for $\epsilon=0$ but is uniformly unstable for all $\epsilon>0$.

The paper is outlined as follows. In Section~\ref{sec:exmpl} we
construct a simple example that shows how arbitrarily small stable
singular perturbations can effect qualitative changes. From
Section~\ref{sec:gr_sl} onward we focus on the case of periodic orbits
with a short region of sliding: this occurs close to so-called
\emph{grazing-sliding} bifurcations. First, in Section~\ref{sec:gr_sl}
we present a numerical observation for a minimal $(2+1)$-dimensional
example based on the Hopf normal form. The example shows that,
depending on the geometry of the problem, sliding may persist or
not. In Section~\ref{sec:gs-sgperturb} we formulate a (still rather
crude) general result about the persistence or destruction of sliding
near grazing. Section~\ref{sec:study} presents the
$\epsilon$-expansions of the local return maps for the minimal Hopf
example.  Finally in Section~\ref{sec:conc} we draw some initial
conclusions and speculate along which lines we hope to generalize our
results beyond the minimal example.

\section{Simple example: destabilization of pseudo-equilibria}
\label{sec:exmpl}

Let us start with the simplest possible example of a Filippov
slow-fast system. This example demonstrates the mechanism by which
stable singular perturbations can destroy stable sliding in Filippov
systems.  Consider the system
\begin{align}
  \label{eq:peqslow}
  \dot x&=-\sgn[\theta x+(1-\theta)y]\\
  \epsilon \dot y&=x-y.\label{eq:peqfast}
\end{align}
In the quasi-static limit $\epsilon=0$ system \eqref{eq:peqslow} and
\eqref{eq:peqfast} has the pseudo-equilibrium $x=y=0$ which is
exponentially stable (even attracting in finite time). We recall that
by a pseudo-equilibrium we mean an equilibrium of \eqref{eq:peqslow}
and \eqref{eq:peqfast} that lies on the switching surface (a
pseudo-equilibrium is typically not an equilibrium of any of the two
flows near the surface). Consider the dynamics of the system
\eqref{eq:peqslow} and \eqref{eq:peqfast} in the neighborhood of this
pseudo-equilibrium for non-zero $\varepsilon.$

Let us express \eqref{eq:peqslow} and \eqref{eq:peqfast} in the
general form given by \eqref{eq:filip}, where
\begin{align*}
f_-
\begin{pmatrix}
  x\\ y
\end{pmatrix}
&=
\begin{bmatrix}
  1\\ \frac{\displaystyle 1}{\displaystyle \epsilon}(x - y)
\end{bmatrix}\mbox{,}
& f_+ \begin{pmatrix}
  x\\ y
\end{pmatrix}
&=
    \begin{bmatrix}
       -1\\ \frac{\displaystyle
      1}{\displaystyle\varepsilon}(x - y)
    \end{bmatrix}
    \mbox{,}
\end{align*}
and the function $h(x,\,y)$ that defines the switching manifold
$\mathcal{H}_s$ is given by
\begin{displaymath}
  h(x,\,y) = \theta x + (1-\theta)y\mbox{.}
\end{displaymath}
The vector normal to $\mathcal{H}_s,$ is $\partial h =
[\theta,\,\,1-\theta]$. The conditions for the existence of an
attracting sliding region read (see \cite{BBCK07})
\begin{equation}
\partial h f_- > 0\mbox{\quad and\quad } \partial h f_+ < 0\mbox{,}
\label{eq:atSl}
\end{equation}
which gives for our example
\begin{align}
  \theta + \frac{1-\theta}{\varepsilon}(x-y)&>0\mbox{,}&
 -\theta + \frac{1-\theta}{\varepsilon}(x-y)&<0\mbox{.}
\label{eq:Cond}
\end{align}
A subset of the switching manifold such that the vector fields $f_-$
and $f_+$ both point towards the switching surface along this subset
(see Figure~\ref{fig:slidesketch}(b)) is called an \emph{attracting
  sliding region}. A \emph{repelling sliding region} is a subset of
the switching manifold such that the vector fields $f_-$ and $f_+$
both point away from the switching surface along this subset.

\begin{figure}[t]
  \centering
  \includegraphics[width=\textwidth]{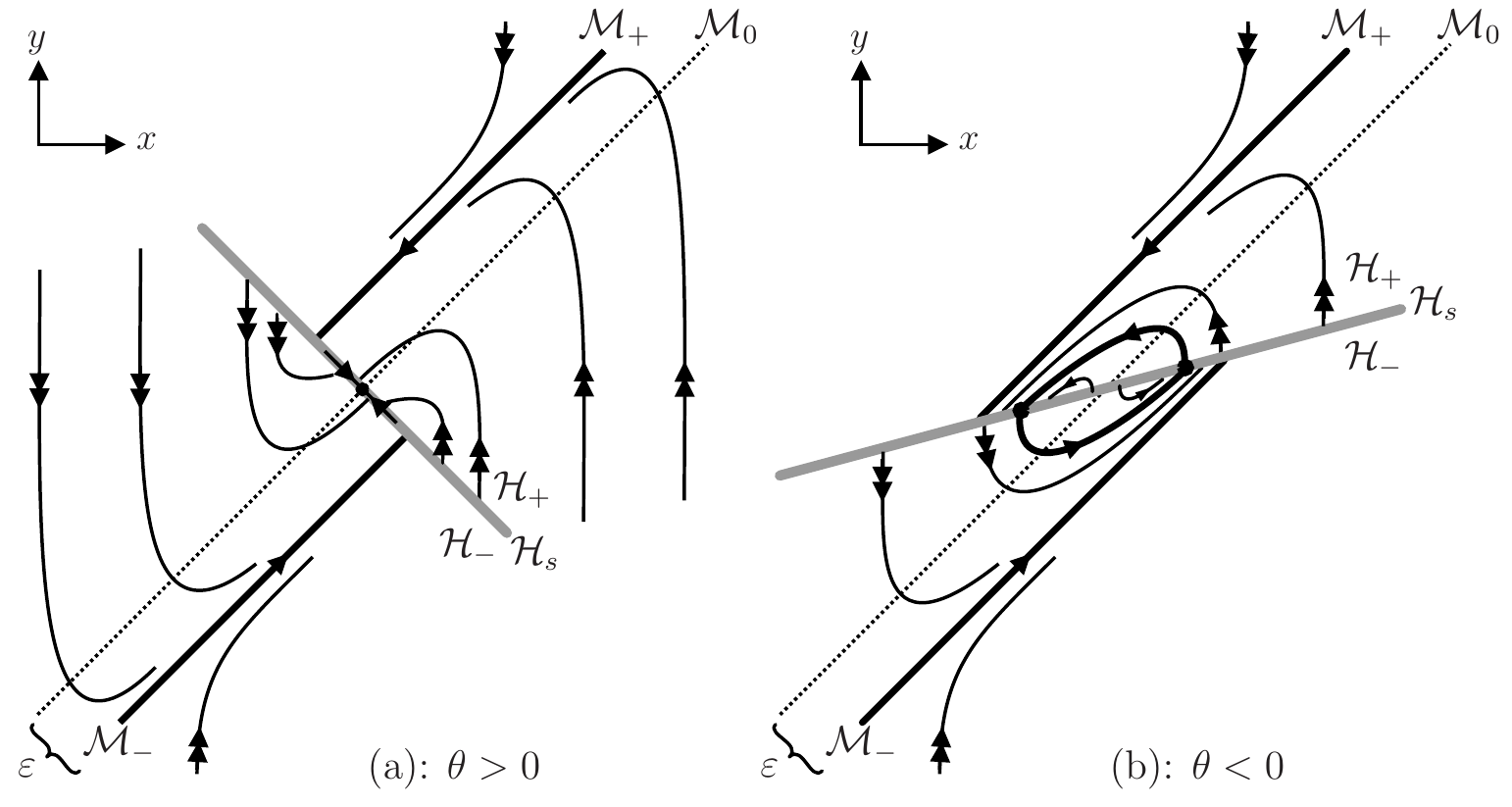}
  \caption{Two cases for
    system~\eqref{eq:peqslow},\,\eqref{eq:peqfast}. The
    pseudo-equilibrium either persists (case (a): $\theta>0$), located
    on the sliding manifold $\mathcal{H}_s=\{\theta
    x+(1-\theta)y=0\}$, together with a small region where the system
    slides, or (case (b): $\theta<0$) it gets perturbed into a small
    periodic orbit and the sliding region near the origin is
    repelling. The manifold $\mathcal{M}_0=\{y=x\}$ is the slow
    manifold, on which the dynamics for $\epsilon=0$ lives. The
    manifolds $\mathcal{M}_\pm=\{y=x\pm\epsilon\}$ are perturbations
    of $\mathcal{M}_0$. Each is invariant with respect to the ODE on
    its side of the switching manifold, containing no rapidly decaying
    dynamics.}
  \label{fig:cases}
\end{figure}

For $\theta\neq1$ the switching manifold $\mathcal{H}_s$ can be
parameterized by $x$, thus, the condition \eqref{eq:Cond} simplifies to
$-\epsilon\theta<x<\epsilon \theta$ on $\mathcal{H}_s$. Therefore, if
$\theta>0$ then \eqref{eq:atSl} holds along a segment of the switching
line $\mathcal{H}_s$ in a neighborhood of the origin (see
Figure~\ref{fig:cases}(a)) and the stable pseudo-equilibrium persists
for positive $\epsilon$ and stays at $x=y=0$. Moreover, in a
sufficiently small neighborhood of the origin the trajectories
converge to this pseudo-equilibrium following the sliding flow.

If $\theta<0$ then (\ref{eq:atSl}) does not hold in the neighborhood
of the origin and there is no stable pseudo-equilibrium (there is a
repelling sliding region near the origin) and instead we have a
symmetric exponentially stable periodic orbit around $(0,0)$ switching
back and forth between $f_+$ and $f_-$. The point
\begin{displaymath}
(x_\mathrm{switch},y_\mathrm{switch})=\left(\left(1-\theta^{-1}\right)
y_\mathrm{switch},y_\mathrm{switch}\right)
\end{displaymath}
 where the periodic orbit
switches from $f_-$ to $f_+$ is parametrized by (see also
Figure~\ref{fig:cases}(b)):
\begin{align*}
  %\begin{split}
    \theta=\frac{1}{2}\,
    \frac{T+2\ef^{-T}
      +\ef^{-T}T-2}{1+\ef^{-T}}\mbox{,}\quad
    y_\mathrm{switch}=\epsilon\theta\,\frac{\ef^{-T}-1}{
      \ef^{-T}+1}
  %\end{split}
\end{align*}
where $\epsilon T$ is the traveling time that the orbit spends on each
side.  Thus, for $|\theta|$ of order $1$ ($\theta<0$) the amplitude
and the period of the periodic orbit are of order $\epsilon$; see
Figure~\ref{fig:cases}(b) for the qualitative picture.  Consequently,
the dynamics of the perturbed system ($\epsilon>0$) are close to the
dynamics of the unperturbed system ($\epsilon=0$) only in a very weak
sense compared to the classical results of \cite{F79}: instead of a
stable fixed point we have an exponentially stable periodic orbit of
small amplitude. This chattering phenomenon is similar to what is
observed when the switching occurs with a small delay
\cite{BKW05,FFS02,SFF03,S06c}.  Example
\eqref{eq:peqslow},\,\eqref{eq:peqfast} shows that chatter can also be
introduced by fast dynamics even if the switching is perfectly
instantaneous.

% If $\theta=1$ then the switching does not depend on the fast variable
% $y$ and there exists an attracting sliding region and the origin
% remains a stable pseudo-equilibrium. On the other hand, if $\theta=0$
% then the sliding region is no longer attracting in the neighborhood of
% the origin (for $\theta=0$ neither attracting nor repelling sliding
% regions exist) but the pseudo-equilibrium is still stable. It becomes
% unstable for $\theta<0$. Hence, if we fix $\epsilon$ at a small but
% non-zero value and vary $\theta$ then the pseudo-equilibrium undergoes
% a discontinuity induced bifurcation at $\theta=0$ (considering
% $\theta$ as a bifurcation parameter; see \cite{KRG03} for bifurcations
% of pseudo-equilibria).  This discontinuity induced bifurcation is
% linked with the fact that the sliding region changes from attracting
% to repelling under the variation of $\theta$ through $0$.
The example also shows that singular perturbations may alter the
defining quantities \eqref{eq:atSl} of the sliding region by order $1$
uniformly for $\epsilon\to0$ and, thus, also change the dynamics
qualitatively (in this example from exponentially stable
pseudo-equilibrium to exponentially stable periodic orbit without
sliding). This will be further highlighted in Section~\ref{sec:study}
where a periodic orbit in a singularly perturbed Filippov system is
analyzed in detail.

Figure~\ref{fig:cases} shows the arrangement of invariant manifolds
and the switching manifold for the example
\eqref{eq:peqslow},\,\eqref{eq:peqfast}. For each of the two flows
there exists an invariant manifold $\mathcal{M}_\pm$ consisting of all
trajectories (in the 2D-example just one trajectory) without rapidly
decaying part. These manifolds $\mathcal{M}_\pm$ are both
$O(\epsilon)$-perturbations of the so-called \emph{slow} manifold
$\mathcal{M}_0$, defined by setting $\epsilon=0$ in the equation
\eqref{eq:peqfast} for the fast dynamics. In the example the manifolds
are $\mathcal{M}_\pm=\{y=x\pm\epsilon\}$ and $\mathcal{M}_0=\{x=y\}$.
Every trajectory that spends a time of order $1$ in the subdomain
$\mathcal{H}_+=\{(x,y): h(x,y)\geq0\}$ ends up exponentially close to
the invariant manifold $\mathcal{M}_+$: its end point has a distance
of order $\exp(-c/\epsilon)$ from $\mathcal{M}_+$ (similarly for
$\mathcal{M}_-$ in $\mathcal{H}_-=\{(x,y): h(x,y)<0\}$).  The example
shows that the invariant manifolds $\mathcal{M}_+$ and $\mathcal{M}_-$
typically differ by a term of order $\epsilon$. This implies that any
trajectory that switches back and forth between $\mathcal{H}_+$ and
$\mathcal{H}_-$ shows a small \emph{boundary layer} immediately after
switching: after switching the trajectory has a short time interval
during which it relaxes to the invariant manifold of the other
flow. These small boundary layers occur also after switching to
sliding and have been studied already in \cite{F00,F02,F02a}. In
example \eqref{eq:peqslow} for $\theta<0$ both parts of the periodic
orbit are still in the boundary layer.

\section{Grazing-sliding in singularly perturbed Filippov systems ---
  illustration}
\label{sec:gr_sl}

In order to understand the effect of stable singular perturbations on
periodic orbits with sliding we focus on periodic orbits with a
\emph{short} sliding segment. One common scenario for periodic orbits
with short sliding segments is the so-called \emph{grazing-sliding}
event; see \cite{BBCK07} for a classification of discontinuity-induced
bifurcations. % It means that
% in the point $x_0=x(t_0)$ not only the switching function $h$ is zero
% but also its time derivative.  Consequently, the periodic orbit $x(t)$
% touches (or grazes) the switching manifold $\{h=0\}$ in $t_0$.
This phenomenon is a \emph{codimension-one} event, which can be
observed generically if the system depends on one additional
parameter.

\subsection{Grazing-sliding --- illustration for minimal example}
\label{sec:gs-min}
\begin{figure}[t]
  \centering
  \includegraphics[width=\textwidth]{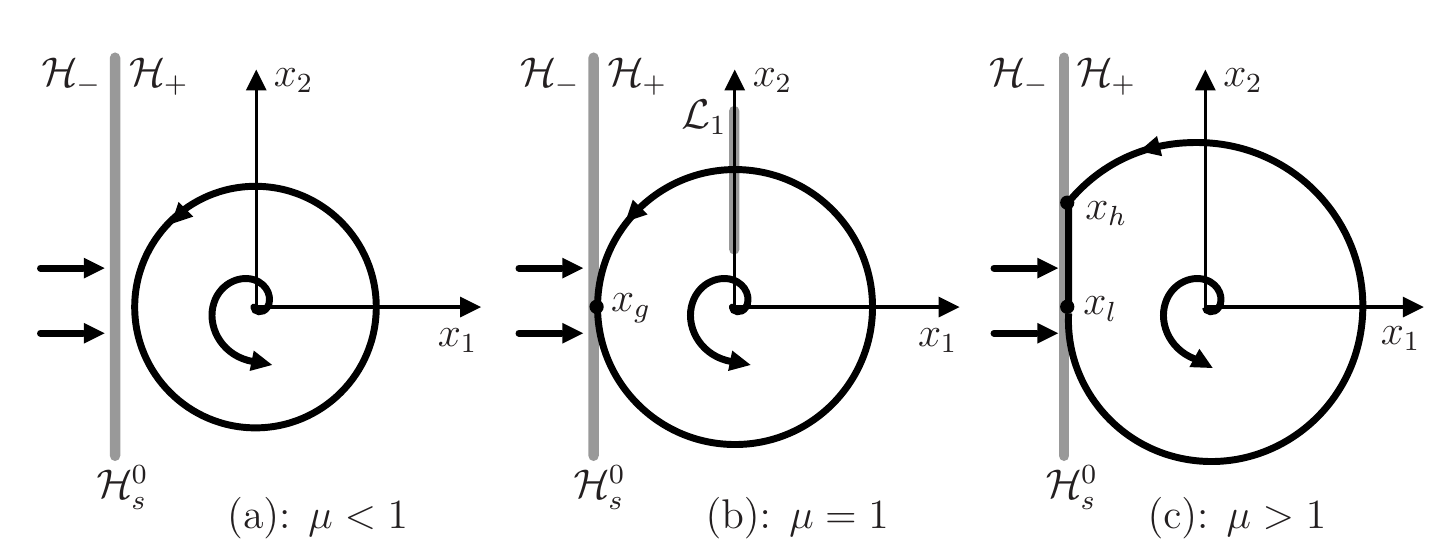}
  \caption{Grazing-sliding scenario for minimal example
    \eqref{eq:red}. (a): the periodic orbit lies entirely in
    $\mathcal{H}_+$; (b): the periodic orbit touches the switching
    line $\{x_1=-1\}$ at $x_g$; (c): the periodic orbit has a small
    sliding segment from $x_h$ to $x_l$. The line $\mathcal{L}_1$ is the
    Poincar{\'e} section chosen for the return map $M_1$ in
    \eqref{eq:locret:graz}.}
  \label{fig:grazingsketch}
\end{figure}

A minimal example for the grazing-sliding event, the Hopf normal form
for $f_+$ combined with a constant vector field for $f_-$, is
\begin{equation}
  \dot x =
  \begin{cases}
   \begin{bmatrix}
     \mu x_1 -\omega x_2 -(x_1^2+x_2^2)x_1\\
     \omega x_1+\mu x_2 -(x_1^2+x_2^2)x_2
   \end{bmatrix}   &\mbox{if\ $x_1\geq-1$}\\
    \begin{bmatrix}
      1\\ 0
    \end{bmatrix} &\mbox{if\ $x_1<-1$,}
  \end{cases}\label{eq:red}
\end{equation}
where we keep $\omega\neq0$ fixed and vary $\mu$ (which is always
greater than $0$) from below $1$ to above $1$ (see
Figure~\ref{fig:grazingsketch}). The switching line for this example is
\begin{displaymath}
  h^0(x)=x_1+1\mbox{,}
\end{displaymath}
and the flow in $\mathcal{H}_+=\{h^0\geq0\}$ has a unique stable
periodic orbit as shown in Figure~\ref{fig:grazingsketch}(a).
In fact, the flow in $\mathcal{H}_+$ can be decomposed into a pair of
uncoupled equations for the polar coordinates $r$ and $\phi$ of
$x\in\R^2$ ($x=r\cdot(\cos\phi,\sin\phi)$):
\begin{align*}
  \dot r&=\ \mu r-r^3\mbox{,}&
  \dot\phi&=\ \omega\mbox{,}
\end{align*}
and the periodic orbit is a circle around the origin corresponding to
$r=\sqrt{\mu}$, $\phi=\omega t$. This periodic orbit is also a
periodic orbit of the Filippov system~\eqref{eq:red} for $\mu<1$.
When we change $\mu$ to values larger than $1$ the periodic orbit of
the Filippov system \eqref{eq:red} acquires a sliding segment starting
from some point $x_h=[-1,x_{h,2}]^T$ ($x_{2,h}>0$) and ending at some
point $x_l=[-1,x_{l,2}]^T$ ($x_{l,2}<0$ is defined by the condition
$\partial hf_+(x_l)=0$), making the overall orbit piecewise smooth (shown in
Figure~\ref{fig:grazingsketch}(c)): it has a corner at $x_h$ and its
second derivative is discontinuous at $x_l$.  The periodic orbit is
grazing the switching manifold $\mathcal{H}_s^0=\{h^0=0\}$ for $\mu=1$
as shown in Figure~\ref{fig:grazingsketch}(b) in a point $x_g$, which
satisfies several conditions simultaneously:
\begin{enumerate}
\item The point $x_g$ lies on the stable periodic orbit of the flow
  in $\mathcal{H}_+$ and on the switching manifold $\mathcal{H}_s^0$.
\item The flow in $\mathcal{H}_+$ is quadratically tangent to the
  switching manifold ($\dot h=0$ and $\ddot h>0$ in $x_g$ along
  trajectories following $f_+$), and
\item the flow in $\mathcal{H}_-$ points toward the
  switching manifold in $x_g$.
\end{enumerate}
The dynamics near the periodic orbit for $\mu\approx1$ is described
completely by its \emph{local return map} (also called
\emph{Poincar{\'e} map}). This map is defined by following the flow
from a small cross-section transversal to the periodic orbit back to
itself. We choose the local return map $M_\mu$ to the section
$\mathcal{L}_1=\{x: x_1=0, x_2>0\}$ (see
Figure~\ref{fig:grazingsketch}(b)), which has for $\mu=1$ the form
\begin{equation}
  \label{eq:locret:graz}
  M_1 x_2=1+
  \begin{cases}
    \ef^{-4\pi/\omega} [x_2-1] + O(|x_2-1|^2)& \mbox{\ if $x_2<1$}\\
    0 & \mbox{\ if $x_2\geq1$.}
  \end{cases}
\end{equation}
Thus, the map $M_1$ is piecewise asymptotically linear in its fixed
point $x_2=1$ corresponding to the periodic orbit. For $\mu>1$ the map
$M_\mu$ is constant near its fixed point. In summary, in system
\eqref{eq:red} the only effect of the grazing under variation of $\mu$
is that the periodic orbit changes its shape, and that its Floquet
multiplier jumps from $\exp(-4\pi \mu/\omega)$ to $0$ at the grazing
parameter $\mu=1$. The periodic orbit is stable for all $\mu\approx
1$.

\subsection{Singular perturbation of grazing-sliding --- observations
  for minimal example}
\label{sec:gs-destruct-min}
Now let us consider a stable singular perturbation of system
\eqref{eq:red}. We will couple \eqref{eq:red} with a one-dimensional stable
fast subsystem for a fast variable $y$ where the coupling also occurs
in the switching function. Thus we obtain
\begin{align}
  \dot x &=\
  \begin{cases}
   \begin{bmatrix}
     \mu x_1 -\omega x_2 -(x_1^2+x_2^2)x_1\\
     \omega x_1+ \mu x_2 -(x_1^2+x_2^2)x_2
   \end{bmatrix}   &\mbox{if\ $\theta x_1+(1-\theta)y\geq-1$}\\
    \begin{bmatrix}
      1\\ 0
    \end{bmatrix} &\mbox{if\ $\theta x_1+(1-\theta)y<-1$}
  \end{cases}\label{eq:exslow}\\
  \epsilon\dot y&=\epsilon\left[\mu x_1 -\omega x_2
    -(x_1^2+x_2^2)x_1\right]+x_1-y.\label{eq:exfast}
\end{align}
The quasi-static approximation replaces $y$ by
$y_0(x,\mu)=x_1$ ($\epsilon=0$ in \eqref{eq:exfast}). Thus, the
reduced model is \eqref{eq:red} which is identical with the slow part
\eqref{eq:exslow} of the singularly perturbed system except that the
switching function depends now in part on the fast variable $y$ for
$\theta\neq1$. Namely
\begin{displaymath}
  h(x,y,\epsilon)=\theta x_1+(1-\theta)y+1\mbox{.}
\end{displaymath}

The question is: how does the perturbation from the reduced system
\eqref{eq:red} to the slow-fast system
\eqref{eq:exslow},\,\eqref{eq:exfast} affect the local return map
$M_1$ of the periodic orbit at $\mu=1$? For $\mu=1$ the grazing
periodic orbit drawn in Figure~\ref{fig:grazingsketch}(b) exists also
in the slow-fast system: its grazing point is now $p_g=(-1,0,-1)$, and
on the orbit the fast variable is given by $y(t)=x_1(t)$. We make the
following observation (see also Figure~\ref{fig:retmaps}):
\begin{figure}[t]
  \centering
  \subfigure[$\theta=-0.5$]{\includegraphics%
    [width=0.48\columnwidth]{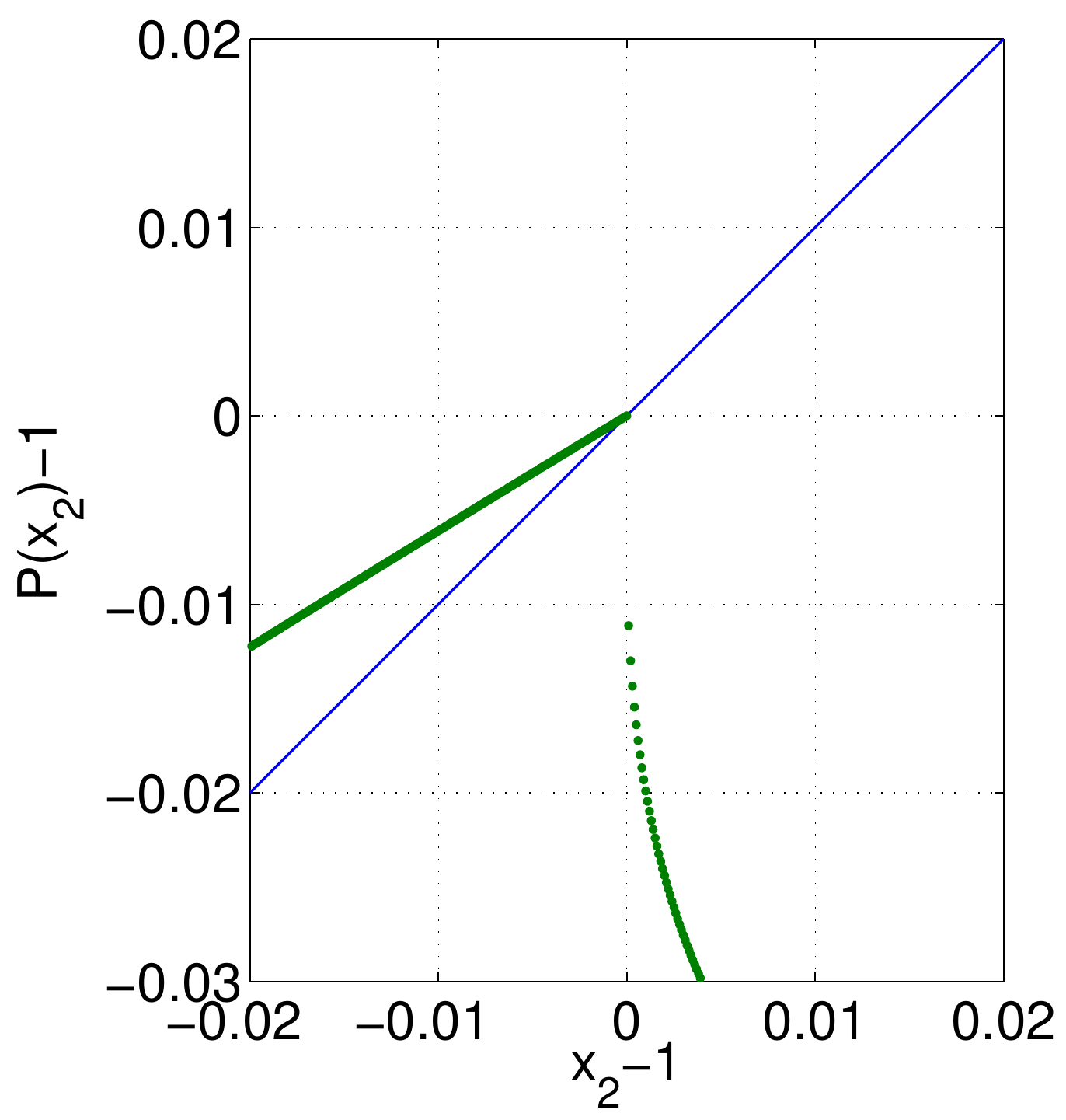}}
  \subfigure[$\theta=0.5$]{\includegraphics%
    [width=0.48\columnwidth]{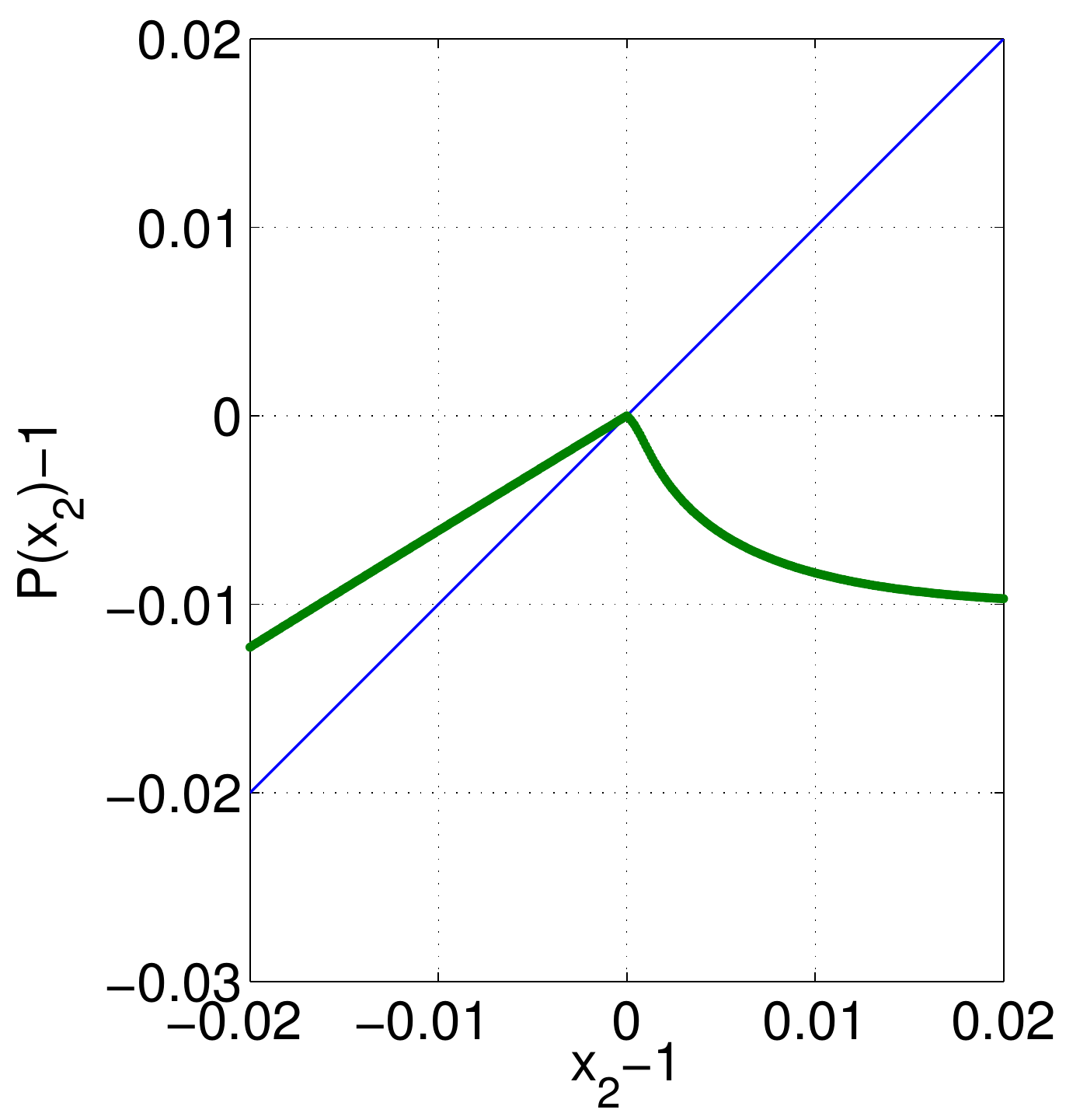}}
  \caption{Approximate local return map to $\mathcal{L}_1$, $P$, for
    $\epsilon=0.01$, $\mu=1$, $\omega=8\pi$ obtained numerically as
    the first return to $\mathcal{L}_1$ (neglecting the distance of
    the return point to the slow invariant manifold
    $\mathcal{M}_+=\{(x,y); x_1=y\}$ of the flow in
    $\mathcal{H}_+$). Intersections with the diagonal indicate fixed
    points corresponding to period-one orbits.}
  \label{fig:retmaps}
\end{figure}
\begin{observation}[Perturbation of local return map near grazing]\ \\
  The perturbed return map for $\epsilon>0$ is a one-dimensional
  return map $P$ to the line $\mathcal{L}_1=\{(x,y); x_1=y=1\}$ if we
  ignore terms of order $\exp(-c/\epsilon)$.

  If $\theta<0$ then the map $P$ is discontinuous for $\epsilon>0$,
  and the discontinuity is of size $O(\epsilon)$.

  If $\theta>0$ then the return map $P$ is continuous also for
  $\epsilon>0$ but the slope near the point $p_g$ is of order $1$
  uniformly for $\epsilon\to0$ and $x_2 \geq 1$.
\end{observation}
Figure~\ref{fig:retmaps} shows how the return map $M_1$ of the grazing
periodic orbit is perturbed for positive $\epsilon$, and
$\theta<0$ (Figure~\ref{fig:retmaps}(a)) and $\theta>0$
(Figure~\ref{fig:retmaps}(b)). The grazing point $p_g$ corresponds to
$x_2=\mu=1$. For $x_2<1$ the return map $P$ is identical to $M_1$. We
observe that for $\theta<0$ the return map becomes discontinuous at
$x_2=1$ and that the size of the jump is of order $\epsilon$. Also, the
limiting slope for $x_2>1$ appears to be $-\infty$. For $\theta>0$ we
note that the limiting slope for $x_2\to+1$ is not of order
$\epsilon$. The slope is negative and of order $1$. For comparison,
the map $M_1$ is identical to $1$ for $x_2>1$. The change of shape of
the graph of the return map from $M_1$ to $P$ has consequences for the
dynamics near the grazing orbit. For $\theta<0$ the period-one orbit
disappears as $\mu$ passes through $1$ from below and we observe
an inverted period-adding cascade \cite{DB09}. For $\theta>0$ the
period-one orbit persists but may change its stability depending on
the slope of the graph in Figure~\ref{fig:retmaps}(b) for
$x_2\to+1$. The local attractor (if any) for $\mu>1$ also depends on
this limiting slope (see \cite{BBCK07} for a recent textbook overview
of possible scenarios in piecewise locally linear maps).

\section{Singular perturbation of grazing-sliding events}
\label{sec:gs-sgperturb}

This section gives a criterion that helps to distinguish whether a given periodic orbit
that grazes the switching surface in a general slow-fast Filippov system
undergoes a bifurcation as shown in Figure~\ref{fig:retmaps}(a) (that is, the local return map is discontinuous) or if we have a grazing-sliding case shown in
Figure~\ref{fig:retmaps}(b) (the local return map is continuous). Let
us assume that the reduced Filippov system~\eqref{eq:filipred} depends
on a parameter $\mu$, and that \eqref{eq:filipred} has a family of
periodic orbits $x(t;\mu)$. We assume without loss of generality that
the periodic orbit $x(t;\mu)$ has no sliding segment near $t=0$ for
$\mu<0$, and it lies entirely in the region $\mathcal{H}_+=\{x:
h(x,y_0(x),0)\geq0\}$ locally near $t=0$ (see
Figure~\ref{fig:grazingsketch}(a)). The precise characterization for a
grazing-sliding event for the family $x(t;\mu)$ at a parameter value
$\mu_0$ ($\mu_0=0$ without loss of generality) can be given by using
the function $\chi_0:(t,\mu)\mapsto h(x(t;\mu),y_0(x(t;\mu)),0)$. The
function $\chi_0$ gives the value of $h$ depending on time along the
periodic orbit.  The family $x(t;\mu)$ grazes at $\mu=0$ if there
exists a time $t_0$ near $0$ such that
\begin{enumerate}
\item\label{cond:sw} $\chi_0(t_0,0)=0$ ($x(t_0;0)$ is on the
  switching manifold $\mathcal{H}_s$),
\item\label{cond:graz} $\partial_1\chi_0(t_0,0)=0$ (the orbit grazes
  $\mathcal{H}_s$ in $t_0$),
\item\label{cond:quad} $\partial_1^2\chi_0(t_0,0)>0$ (the grazing is
  quadratic),
\item\label{cond:genparam} $\partial_2\chi_0(t_0,0)<0$ (the parameter
  $\mu$ unfolds the grazing),
\item\label{cond:slid} $[\partial_1 h+\partial_2 h\,\partial_2
  g^{-1}\partial_1g]f_->0$ in $(x(t_0;0),y_0(x(t_0;0)),0)$ (the
  reduced vector field in $\mathcal{H}_-=\{x: h(x,y_0(x),0)<0\}$ points
  toward switching manifold).
\end{enumerate}
(The functions $h$, $f_\pm$ and $y_0$ may all depend on $\mu$ in this
list.)  Note that this is a codimension-one event because we have two
equality conditions (\ref{cond:sw}) and (\ref{cond:graz}) and can
adjust the time $t_0$. The conditions \ref{cond:sw}--\ref{cond:slid}
guarantee that the periodic orbit and the switching manifold interact
as shown in Figure~\ref{fig:grazingsketch}(b). See also \cite{BBCK07}
for a detailed discussion of grazing-sliding events for periodic
orbits.

Furthermore, let us assume that the periodic orbit $x(t;\mu)$ is
entirely in $\mathcal{H}_+$ for $\mu<0$ (not only locally near $t_0$).
Then the standard singular perturbation theory of \cite{F79}
guarantees that for sufficiently small $\epsilon>0$ the full system
\eqref{eq:sgfilipslow},\,\eqref{eq:sgfilipfast} also has a family of
periodic orbits $(x,y)(t;\mu,\epsilon)$ which depends smoothly on
$\mu$ up to some parameter value $\mu_\epsilon=O(\epsilon)$ where it
grazes quadratically at a point
$(x_\epsilon,y_\epsilon)=(x,y)(t_0;\mu_\epsilon,\epsilon)$. We can
keep the time point of the grazing equal to $t_0$ independent of
$\epsilon$ without loss of generality by shifting the time in the
family of periodic orbits because the slow-fast system
\eqref{eq:sgfilipslow},\,\eqref{eq:sgfilipfast} is autonomous. This
means that the function $\chi_\epsilon:(t,\mu)\mapsto
h(x(t;\mu,\epsilon),y(t;\mu,\epsilon),\epsilon)$ also satisfies the
first four conditions \ref{cond:sw}--\ref{cond:genparam} for
$\mu=\mu_\epsilon$ and $t_0$. This is a simple persistence argument
because only the smooth vector field $f_+$ is involved.  However, in
general the equivalent of condition \ref{cond:slid} (that the vector
field on the other side points toward the switching manifold) is not
necessarily satisfied for the full system
\eqref{eq:sgfilipslow},\,\eqref{eq:sgfilipfast} even though it is
satisfied for the reduced system.

The following lemma states how this affects the local return map of
the periodic orbit at $\mu_\epsilon$:
\begin{lemma}[Grazing-sliding of periodic orbits]\label{thm:gen}
  Suppose that the reduced system \eqref{eq:filipred} depends on a
  parameter $\mu$ and has a grazing-sliding periodic orbit, lying
  entirely in $\mathcal{H}_+$, and satisfying the conditions
  \ref{cond:sw}--\ref{cond:slid} for the parameter value $\mu=0$ at
  $x_0=x(t_0;0)$. % Furthermore, assume that the
%   periodic orbit $x(t;0)$ does not have any multipliers on the unit
%   circle (except for one trivial multiplier equal to $1$) for small
%   $\mu$.

  Then for sufficiently small $\epsilon>0$ the family of periodic
  orbits in the full system
  \eqref{eq:sgfilipslow},\,\eqref{eq:sgfilipfast} also grazes for a
  nearby parameter value $\mu_\epsilon=O(\epsilon)$ and we have two
  generic cases.
  \begin{description}
  \item[Case 1] If $\partial_1 h f_--\partial_1h f_+>0$ in
    $(x_0,y_0(x_0),0)$ then for sufficiently small $\epsilon>0$ the
    local return map of the grazing periodic orbit is continuous (and
    piecewise smooth).
  \item[Case 2] If $\partial_1 hf_--\partial_1 hf_+<0$ in
      $(x_0,y_0(x_0),0)$ then for sufficiently small
    $\epsilon>0$ % the sliding condition is not satisfied in
%     $\mathcal{H}_-$ near $x_0$.  Hence,
    % if the matrix
%     \begin{displaymath}
%       \begin{bmatrix}
%         f_- & f_+\\[-1ex] g & g
%       \end{bmatrix}
%     \end{displaymath}
%     has full rank in $(x_0,y_0(x_0),0)$ then
    the local return map of the grazing periodic orbit is
    discontinuous in the fixed point corresponding to the grazing
    periodic orbit.
  \end{description}
\end{lemma}
In our set-up the local return map does not depend on the choice of
the cross-section as long as the cross-section is not taken at $t_0$
(return maps to different cross-sections are equivalent and the
coordinate change is a diffeomorphism obtained by following the flow
in $\mathcal{H}_+$).

As the minimal example \eqref{eq:filipred} shows, a typical feature of
the grazing-sliding event is that the local return map along the
periodic orbit is piecewise smooth in its fixed point corresponding to
the periodic orbit $x(\cdot;\mu)$ for the parameter $\mu=\mu_\epsilon$
at which grazing occurs.  More precisely, the return map is smooth
(with a uniformly bounded derivative) everywhere except along a
manifold containing the fixed point. At this manifold the map is only
(Lipschitz) continuous. The lemma states that in the second case the
return map of the grazing orbit in the singularly perturbed system
develops a discontinuity for positive $\epsilon$ at this manifold. The
appearance of a discontinuity is guaranteed because we assume that the
grazing orbit lies entirely in one domain, always following the flow
$(f_+,g)$ (see Appendix~\ref{sec:lemmaproof} for details). In
Section~\ref{sec:study} we show that for the minimal example
\eqref{eq:exslow},\,\eqref{eq:exfast} the size of the discontinuity
can be expected to be of order $\epsilon$.

The quantity distinguishing case~1 and case~2 can be evaluated without
knowledge of the right-hand-side $g$ of the fast dynamical subsystem
\eqref{eq:sgfilipfast}. This means that one can check if sliding
persists under stable singular perturbations of a term $y_0(x)$ in the
reduced model without having to know how this perturbation looks like.

The two cases are distinguished by checking if the vector field in the
domain $\mathcal{H}_-=\{(x,y): h(x,y,\epsilon)<0\}$ points toward the
switching surface $\mathcal{H}_s$ near the grazing point. It turns out
that in the second case the reason for the discontinuity of the return
map is the presence of a small repelling sliding region. For the first
case the sliding condition is satisfied such that sliding occurs and
the general theory developed in \cite{BBCK07} (Theorem~8.1)
applies. The proof of Lemma~\ref{thm:gen}, given in
Appendix~\ref{sec:lemmaproof}, simply has to check this sliding
condition. Theorem~8.1 of \cite{BBCK07} can also be used to
approximate the one-sided derivatives of the local return map of the
full system \eqref{eq:sgfilipslow},\,\eqref{eq:sgfilipfast} for the
sliding case (case 1) of Lemma~\ref{thm:gen}. We show in
Section~\ref{sec:study} for the minimal example
\eqref{eq:exslow},\,\eqref{eq:exfast} that the limit for
$\epsilon\to0$ for these one-sided derivatives can be different from
the one-sided derivatives obtained for the reduced model.

Lemma~\ref{thm:gen} is rather crude: a more detailed analysis is
required to find out how the return map of the grazing periodic orbit
actually looks like and how its features (for example, the size of the
discontinuity or the one-sided derivatives) depend on $\epsilon$. This
analysis is technical for a general system such as
\eqref{eq:sgfilipslow},\,\eqref{eq:sgfilipfast}, in particular if the
dimension of the fast variable $y$ is larger than $1$. Thus, we
construct the $\epsilon$-expansions for the return map only for the
minimal example \eqref{eq:exslow},\,\eqref{eq:exfast}. The example
allows us to address both cases of Lemma~\ref{thm:gen} by varying its
system parameter $\theta$.  As long as the fast variable is
one-dimensional, the generalization to arbitrary singularly perturbed
systems with a grazing-sliding periodic orbit is straightforward.

\section{Expansion of return maps for the minimal example}
\label{sec:study}
Let us return to the minimal example
\begin{align}
  \dot x &=\
  \begin{cases}
   \begin{bmatrix}
     \mu x_1 -\omega x_2 -(x_1^2+x_2^2)x_1\\
     \omega x_1+ \mu x_2 -(x_1^2+x_2^2)x_2
   \end{bmatrix}   &\mbox{if\ $\theta x_1+(1-\theta)y\geq-1$}\\
    \begin{bmatrix}
      1\\ 0
    \end{bmatrix} &\mbox{if\ $\theta x_1+(1-\theta)y<-1$}
  \end{cases}\label{eq:exslow2}\\
  \epsilon\dot y&=\epsilon\left[\mu x_1 -\omega x_2
    -(x_1^2+x_2^2)x_1\right]+x_1-y\label{eq:exfast2}
\end{align}
where $\omega>0$, $\theta<1$ and $\mu\approx1$. We have chosen the right-hand-side
$g$ of the fast equation \eqref{eq:exfast2} such that for $\epsilon>0$
the subspace $\mathcal{ M}_+=\{(x,y): x_1=y\}$ is the exact slow
invariant manifold of the flow $E_+^t$ (following $\dot x=f_+$,\
$\epsilon\dot y=g$).  That is, any trajectory will converge to
$\mathcal{ M}_+$ with a convergence rate of order $\epsilon^{-1}$ as
long as it stays in the half space $\mathcal{ H}_+=\{(x,y): \theta
x_1+(1-\theta)y\geq-1\}$. The graph of the invariant manifold
$\mathcal{ M}_+$ for all $\mu$ and $\epsilon$ is given by
\begin{equation}
  y_{m,+}(x)=x_1\label{eq:ymp}\mbox{.}
\end{equation}
\begin{figure}[t]
  \centering
  \includegraphics[scale=1]{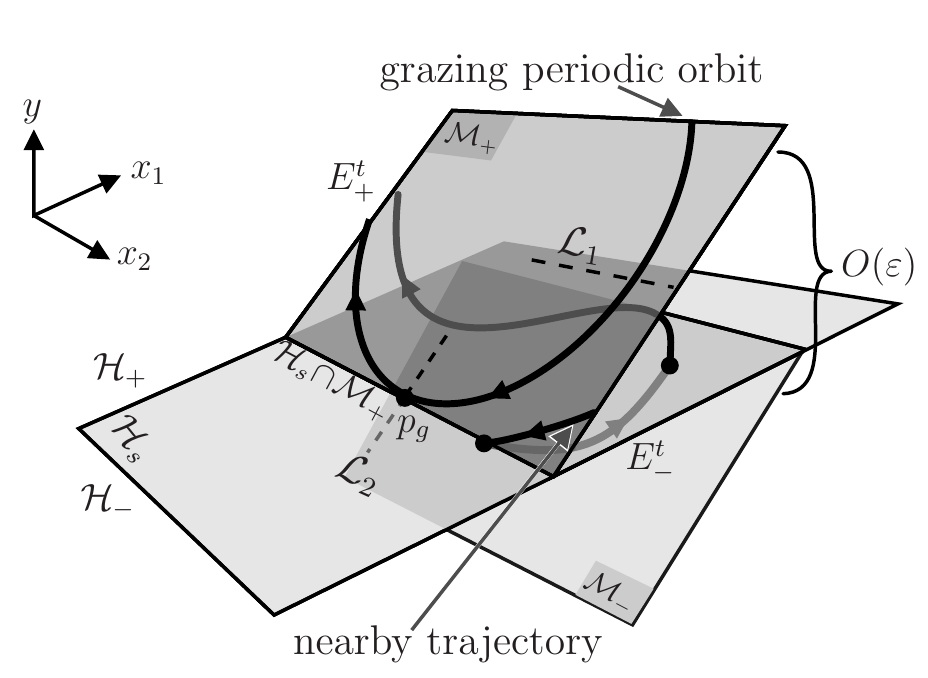}
  \caption{Illustration of the manifolds near the grazing point $p_g$
    for the slow-fast example \eqref{eq:exslow2},\,\eqref{eq:exfast2}
    for $\epsilon>0$, $\theta<0$ and $\mu=1$ (zoom-in near the grazing
    point $p_g$ of the periodic orbit). The grazing periodic orbit
    lies entirely in $\mathcal{H}_+$ (and, thus, in $\mathcal{M}_+$).
    The other invariant manifold $\mathcal{M}_-$ (for $\mathcal{H}_-$)
    has a distance of $O(\epsilon)$ from $\mathcal{M}_+$. The
    illustration shows the grazing periodic orbit and a typical (for
    $\theta<0$) switching trajectory near the periodic orbit,
    switching from $E_+$ to $E_-$, back to $E_+$, and then
    re-approaching $\mathcal{M}_+$.}
  \label{fig:thetalt0mfs}
\end{figure}
Since $f_+$ does not depend on $y$ the \emph{stable fibres} of the
manifold $\mathcal{M}_+$ are: $\mathcal{
  F}_+(x_0)=\{(x,y):x=x_0,\,y\in\R\}$ for all $x_0\in\R^2$ (all points
within a stable fibre converge to each other with a rate $O(\epsilon^{-1})$
following flow $E_+^t$ forward in time).  The graph of the slow
invariant manifold $\mathcal{ M}_-$ of $E_-$ (the flow following $\dot
x=f_-$,\ $\epsilon\dot y=g$) is not known analytically. Its expansion
up to order $\epsilon$ is
\begin{equation}
  \label{eq:ymm}
  y_{m,-}(x,\mu,\epsilon)=x_1
  +\epsilon\left[\mu x_1-\omega x_2-(x_1^2+x_2^2)x_1-1\right]
  +O(\epsilon^2)\mbox{.}
\end{equation}
The expressions for the slow invariant manifolds for $E_+$ and $E_-$,
\eqref{eq:ymp} and \eqref{eq:ymm} differ from each other by a term of
order $\epsilon$; the difference is $\epsilon+O(\epsilon^2)$ for
$x=(-1,0)$, $y=-1$, $\mu=1$, thus, it is non-zero at the grazing point
$p_g=(-1,0,-1)$ at $\mu=1$. Hence, we expect that any
trajectory crossing the switching manifold will show a small boundary
layer; see Figure~\ref{fig:thetalt0mfs} for an illustration of the
invariant manifolds $\mathcal{M}_\pm$ and the switching manifold
$\mathcal{H}_s=\{(x,y): \theta x_1+(1-\theta)y=-1\}$ near the grazing
point at $\mu=1$.

The switching manifold $\mathcal{H}_s$ for the example
\eqref{eq:exslow2},\,\eqref{eq:exfast2} is a two-dimensional plane. Its
intersection with the invariant manifold $\mathcal{ M}_+$ is
\begin{displaymath}
    \mathcal{ H}_s\cap\mathcal{ M}_+=\{(x,y): x_1=y=-1\}\mbox{.}
%     \mathcal{ H}_s\cap\mathcal{ M}_-&=\{(x,y):
%     x_1=-1+\epsilon(1-\theta)[p+\omega x_2-x_2^2]+O(\epsilon^2)\\
%     &\phantom{=\{(x,y):}\ y=-1-\epsilon\theta[p+\omega
%     x_2-x_2^2]+O(\epsilon^2)\}
%   \end{split}
\end{displaymath}

The two cases of Lemma~\ref{thm:gen} correspond to $\theta>0$ (Case 1)
and $\theta<0$ (Case 2). They have to be treated differently and
result in qualitatively different return maps as
Figure~\ref{fig:retmaps} shows.  The return maps for both cases are
superpositions of a smooth global return map (following $E_+$) and a
correction in the vicinity of the grazing, the Poincar{\'e}-section
discontinuity mapping (PDM, \cite{BBCK07}).  In the following we
derive an expression for the return map to the section
$\Sigma_1=\{(x,y): x_1=0,x_2>0,y\in\R\}$ which is away from the
grazing point (see Figure~\ref{fig:thetalt0mfs} for
$\mathcal{L}_1=\Sigma_1\cap\mathcal{M}_+$). More specifically
we consider the
local return map to a small neighborhood $\mathcal{ U}\subset
\Sigma_1$ of the point $(x,y)^T=(0,1,0)^T$, the fixed point
at $\mu=1$ corresponding to the grazing periodic orbit. Any
trajectory starting in $\mathcal{ U}$ spends a time of approximately
$\pi/(2\omega)$ following $E_+$ before it reaches the vicinity of the
switching manifold $\mathcal{ H}_s$ in a small neighborhood $\mathcal{
  V}$ of the point $p_g=(-1,0,-1)^T$. During this time the difference
between $y$ and $x_1$ (which is the distance to the slow invariant
manifold $\mathcal{M}_+$) decays such that
$|y-x_1|\sim\exp(-\pi/(2\omega\epsilon))$ when the trajectory reaches
$\mathcal{ V}$, which is beyond all orders of $\epsilon$. A trajectory
leaving $\mathcal{ V}$ follows $E_+$ for a time of approximately
$3\pi/(2\omega)$ until it reaches $\Sigma_1$. Again, after this time
the $y$-component and the $x_1$-component of the trajectory will be
$\exp(-2\pi/(3\omega\epsilon))$ close to each other. This means that
the return map $P$ is a one-dimensional map from the line
$\mathcal{L}_1=\Sigma_1\cap\mathcal{M}_+=\{(x,y): x_1=y=0,x_2>0\}$
back to itself if we ignore terms of order $\exp(-c/\epsilon)$ where
$c>0$ is of order $1$ (the return line $\mathcal{L}_1$ is also
indicated by a dashed line in Figure~\ref{fig:thetalt0mfs}). The
overall return map $P$ is a composition of four maps.  Calling the
time derivative of $h$ with respect to each of the flows as $h'_\pm$,
respectively,
\begin{displaymath}
    h'_\pm(x,y)=\ \partial h(x,y)\frac{d}{dt}
    E_\pm^t(x,y)\vert_{t=0}=[\partial_1hf_\pm+
    \epsilon^{-1}\partial_2hg](x,y)\mbox{,}
\end{displaymath}
(dropping the argument $\epsilon$) the maps $P_1$ to $P_4$ act as follows
\begin{enumerate}
\item $P_1$, maps from the line $\mathcal{
    L}_1=\{x_1=y=0,x_2\}\cap\mathcal{ U}$ to the curve $\mathcal{
    L}_2=\{x_1=y,h'_+(x,y)=0\}\cap\mathcal{ V}$ by following the flow
  $E_+$ within the slow invariant manifold $\mathcal{ M}_+$ (see
  Fig.~\ref{fig:thetalt0mfs}),
\item $P_2$, the Poincar{\'e}-section discontinuity mapping from
  $\mathcal{ L}_2$ to the surface $\Sigma=\{(x,y): h'_+(x,y)=0\}\cap
  \mathcal{ V}$, which contains the curve $\mathcal{L}_2$,
\item $P_3$, maps from $\Sigma$ back to $\mathcal{ L}_2$, following the
  projection along the stable fibres of $E_+$:
  $P_3(x_1,x_2,y)=(x_1,x_2,x_1)$,
\item $P_4$, maps from $\mathcal{ L}_2\subset \mathcal{ V}$ back to
  $\mathcal{ L}_1 \subset \mathcal{ U}$ following $E_+$ inside the
  slow invariant manifold $\mathcal{M}_+$.
\end{enumerate}
% The curve $\mathcal{ L}_2$ can be parametrized locally by $\tilde x_1$, that is
% \begin{displaymath}
%   \mathcal{ L}_2=\{(\tilde x,\tilde y):
%   \tilde x_2=\eta(\tilde x_1), \tilde y=\tilde x_1\}
% \end{displaymath}
% for some function $\eta$, because the tangency of the periodic orbit
% at the grazing is quadratic.
% The quantity $\Delta$, defining the
% length of the line $\mathcal{ L}_1$, is (to be chosen) sufficiently
% small but larger than $\epsilon$.
The composition
\begin{displaymath}
P=P_4\circ P_3\circ P_2\circ P_1
\end{displaymath}
maps the line $\mathcal{ L}_1$ back onto itself and it is a
one-dimensional approximation of the true two-dimensional return map
near the grazing periodic orbit up to terms of order
$\exp(-c/\epsilon)$. The map $P$ is conjugate (up to
the diffeomorphism $P_1$) to the map
\begin{displaymath}
  [P_1 \circ P_4]\circ [P_3 \circ P_2]=:
  P_\mathrm{glob}\circ P_\mathrm{DM}\mbox{,}
\end{displaymath}
which maps from $\mathcal{ L}_2$ back to itself. The smooth map
$P_\mathrm{glob}=P_1\circ P_4:\mathcal{L}_2\mapsto\mathcal{L}_2$ is
the global return map around the periodic orbit following the smooth
flow $E_+$. Thus, $P_\mathrm{glob}$ has the fixed point
\begin{equation}
  x_{1,\mathrm{glob}}=-\sqrt{\mu}\label{eq:fpg}
\end{equation}
corresponding to the stable periodic orbit of $E_+$. Its
linearization at the fixed point is stable and is identical to the
non-trivial Floquet multiplier of the periodic orbit of $E_+$, which
is $\exp(-4\pi\mu/\omega)$. This means that the smooth global
return map $P_\mathrm{glob}=P_1\circ P_4$ is described to first order
by
\begin{equation}
  \label{eq:gret}
  P_\mathrm{glob}(x_1)= -\sqrt{\mu}+
  \exp\left(-4\pi\mu/\omega\right)
  \left[x_1-x_{1,\mathrm{glob}}\right]
  +O([x_1- x_{1,\mathrm{glob}}]^2).
\end{equation}
The other part of the return map $P$, the projected discontinuity mapping
\begin{displaymath}
P_\mathrm{DM}=P_3\circ P_2
\end{displaymath}
depends on the sign of the parameter $\theta$ in the switching
function $h$. For each sign of $\theta$ we study a range of $\mu$, $x$
and $y$ near the grazing point $(x,y)=p_g=(-1,0)$, $\mu=1$. The size of the
range depends on $\epsilon$: it is $O(\epsilon)$ for $\theta<0$, and
it is $O(\epsilon^2)$ for $\theta>0$. We take this into account by
coordinates corresponding to a zoom-in into the neighborhood of the
the grazing point $p_g$ (blowing up the small neighborhood to size
$1$).
\begin{lemma}[Discontinuity mappings]\label{thm:pdm}\ \\
  \begin{description}
  \item[Case $\theta<0${\rm:}] Introducing the coordinates $\xi_1$ and $q$ by
\begin{math}
    x_1=-1+\epsilon\xi_1
\end{math} and
\begin{math}
    \mu=1+\epsilon q
\end{math},
the map $P_\mathrm{DM}$ has the form
\begin{equation}
  \label{eq:pdmtlt0}
  P_{\mathrm{DM}}(\xi_1)=
  \begin{cases}
    \xi_1 &\mbox{if $\xi_1\geq0$}\\
    s_0(\theta)+\sqrt{\epsilon}\left[\frac{\omega
        s_0(\theta)}{\theta+s_0(\theta)}\sqrt{-2\xi_1}+
      \frac{1}{\omega}(-2\xi_1)^{3/2}\right]+O(\epsilon)& \mbox{if
      $\xi_1<0$.}
  \end{cases}
\end{equation}
\item[Case $\theta>0${\rm:}]  Introducing the coordinates $\xi_1$ and $q$ by
\begin{math}
    x_1=-1+\epsilon^2\xi_1
\end{math} and
\begin{math}
    \mu=1+\epsilon^2 q
\end{math},
the map $P_\mathrm{DM}$ has the form
\begin{equation}
  \label{eq:pdmtgt0}
  P_{\mathrm{DM}}(\xi_1)=
  \begin{cases}
    \xi_1 &\mbox{if $\xi_1\geq0$}\\
    s(\theta,\omega,\xi_1,\epsilon)\xi_1& \mbox{if $\xi_1<0$}
  \end{cases}
\end{equation}
where the graph $s(\theta,\omega,\xi_1,\epsilon)\xi_1$ is
parametrically defined by the sliding time $t_s>0$. The zero-order
term can be represented as
\begin{equation}
    \label{eq:xi1r}
  \begin{split}
    \xi_1&=\ -\frac{\omega^2\theta^2}{2}\left[\frac{(1-\theta)
        \left[1-\exp(-t_s/\theta)\right]
        +t_s}{\theta+(1-\theta)\exp(-t_s/\theta)}\right]^2\\
    s(\theta,\omega,\xi_1,0)\xi_1&=\ \omega^2(1-\theta)\,
    \frac{\theta-\left[\theta+t_s\right]
      \exp(-t_s/\theta)}{\theta+(1-\theta)\exp(-t_s/\theta)}\mbox{.}
  \end{split}
\end{equation}
\end{description}
\end{lemma}
The quantity $s_0(\theta)$ appearing in \eqref{eq:pdmtlt0} is a
uniformly positive constant (only depending on $\theta$). It is
implicitly defined by the equation \eqref{eq:s0def} in
Appendix~\ref{sec:discont} (see also graph of $s_0(\theta)$ in
Figure~\ref{fig:s0theta}).  The remainder term $O(\epsilon)$ in
\eqref{eq:pdmtlt0} for $\xi_1<0$ may contain small corrections to the
constant and square-root terms in $\xi_1$. However, the form
\eqref{eq:pdmtlt0} guarantees that for sufficiently small values of
the singular perturbation parameter $\epsilon$ the overall return map
$P_\mathrm{glob}\circ P_\mathrm{DM}$ changes from a piecewise linear
map for the reduced model ($\epsilon=0$) to a discontinuous map for
the full model ($\epsilon>0$). The size of the jump at the
discontinuity is of order $\epsilon$ in the original coordinates.
Moreover, the slope of the map next to the discontinuity is infinity
from one side.  Figure~\ref{fig:retmaps}(b) shows a numerically
computed graph of the approximate overall return map $P$ to
$\mathcal{L}_1$ for $\theta=-0.5$, $\epsilon=10^{-2}$ and $\mu=1$ near
the critical value of the coordinate $x_2$. This graph neglects
exponentially small terms (the numerical distance of the return value
to the slow invariant manifold $\mathcal{M}_+$ is of the order
$10^{-10}$ at the return section).  Appendix~\ref{sec:discont} gives
the details of the derivation of the expansion of the map
$P_\mathrm{DM}$ in \eqref{eq:pdmtlt0}.

\begin{figure}[t]
  \centering
  \includegraphics[height=0.35\textwidth]{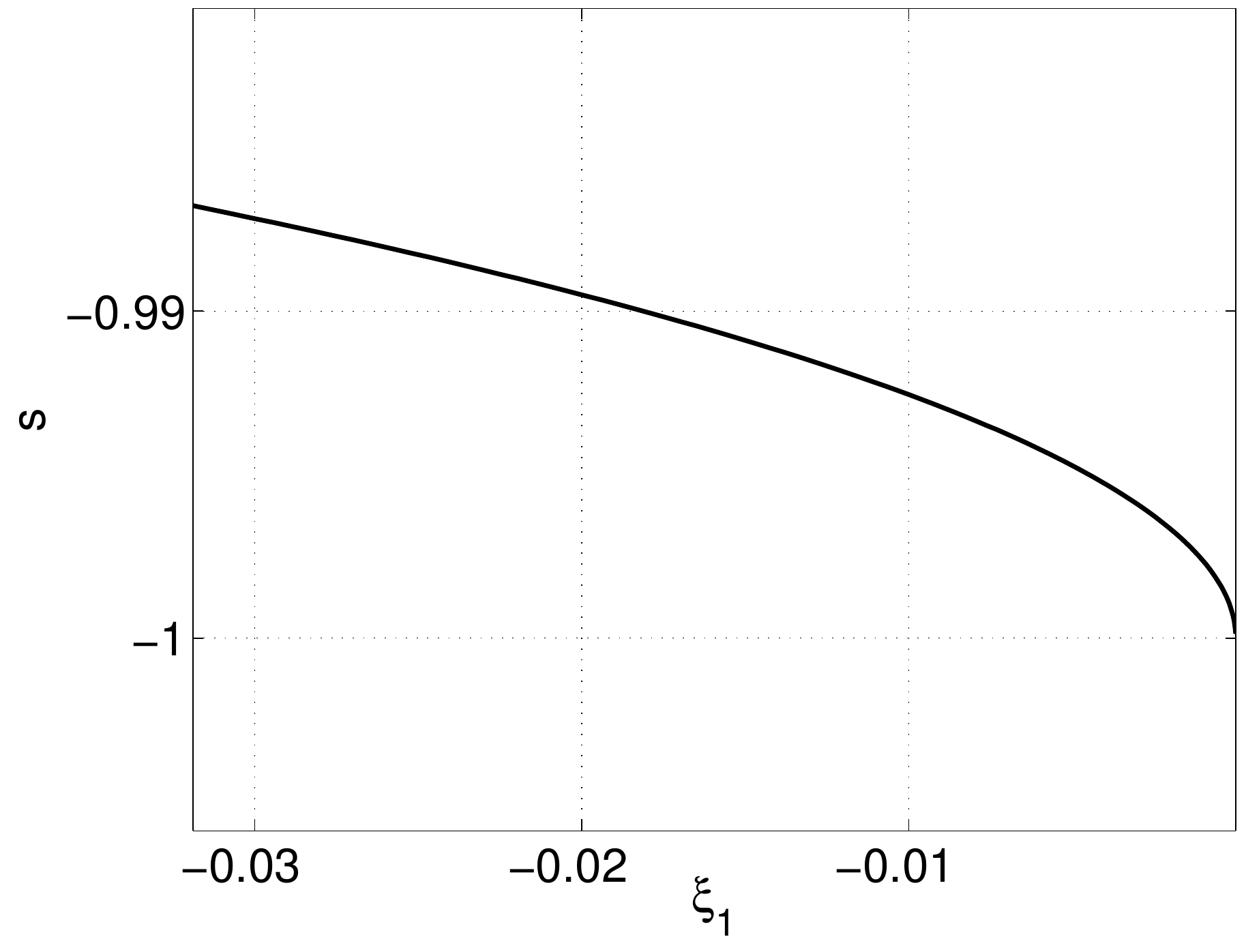}
  \caption{Illustration of the implicitly defined graph
    $s(\theta,\omega,\cdot,0)$, as defined by \eqref{eq:xi1r}
    (parametrized by $t_s$, $\omega=8\pi$, $\theta=0.5$).}
  \label{fig:p:xi12}
\end{figure}
%delta=0.005, epsilon=0.01, omega= 8\pi, theta=0,0.5,-0.5
In \eqref{eq:pdmtgt0} the function $s$ is smooth in all arguments such
that higher order terms do not destroy the continuity of the map
$P_\mathrm{DM}$. Figure~\ref{fig:p:xi12} shows the graph of
$s(0.5,8\pi,\xi_1,0)$.  The value of $s$ at $\xi_1=0$ (that is,
$t_s=0$) is $(\theta-1)/\theta$.  Appendix~\ref{sec:slid} gives the
details of the derivation for the discontinuity mapping
$P_\mathrm{DM}$ in \eqref{eq:pdmtgt0}.  The change of the slope $s$
from its value at $\xi_1=0$ to slope $0$ for $\xi_1\to-\infty$ (as
predicted by the reduced model) occurs over a range of order
$\epsilon^2$ in the original coordinates after grazing.
Figure~\ref{fig:retmaps}(a) shows the approximate overall return map
$P$ to $\mathcal{L}_1$ for $\theta=0.5$, $\epsilon=10^{-2}$ and
$\mu=1$ near the critical value of the coordinate $x_2$.

% \subsection{Dry-friction oscillator revisited}
% \label{sec:dfo2}

\section{Conclusion}
\label{sec:conc}
Stable singular perturbations have a much stronger influence on the
dynamics in Filippov systems than in smooth dynamical systems. We
demonstrate that stable pseudo-equilibria and stable periodic orbits
with sliding do not necessarily persist. We study periodic orbits with
an infinitesimally small sliding segment, that is, close to a
grazing-sliding bifurcation. We found two generic cases depending on
the geometry: the local return map around the grazing periodic orbit
develops a discontinuity if the condition on the existence of an
attracting sliding region is violated. Otherwise, the continuity of
the return map persists but the asymptotic slope may have a change of
order $1$ (uniform for $\epsilon\to0$).

The qualitative change of the local return map induces qualitative
changes to the dynamics on a small scale. A piecewise discontinuous
map with a square-root singularity of the slope on one side of the
discontinuity, as occurs for $\theta<0$ in the minimal example, shows
inverted period-adding cascades of periodic orbits if one varies the
parameter $\mu$ through its critical value \cite{DB09}. The parameter
range where these cascades can be observed is of order $\epsilon$. In
the other case, $\theta>0$, the observed dynamics depends strongly on
the one-sided derivative $s(\theta,0,0)$ defined in
Lemma~\ref{thm:pdm}. It can be chaotic if $s(\theta,0,0)<-1$, which is
possible for small $\theta$. Our analysis is valid on a scale of order
$\epsilon^2$ in phase space and parameter space.

The results of Lemma~\ref{thm:pdm} can be generalized to
higher-dimensional slow-fast systems in a straightforward manner as
long as the dimension of the fast subsystem is $1$. There are,
however, some technical difficulties to generalizing the expressions
of Lemma~\ref{thm:pdm} for higher dimensional fast subsystems
($y\in\R^m$, $m>1$); a trajectory following the dynamics inside the
stable fibres (following a linear stable ODE) may intersect the
switching hyperplane several times. For the expressions
\eqref{eq:pdmtlt0} and \eqref{eq:pdmtgt0} in Lemma~\ref{thm:pdm} we
exploited that this is not the case. In $\R^1$ every trajectory in a
stable linear system approaches the origin in a monotone (increasing
or decreasing) fashion, which is not true in $\R^2$ in the Euclidean
norm. Furthermore, the Poincar{\'e}-section discontinuity mapping is
only implicitly given as the root of a nonlinear equation even for the
minimal example. In general, this implicit expression is determined by
the intersection of a trajectory following a stable linear system with
a hyperplane.

What are the consequences of the small-scale instabilities described
in our paper in practical applications? They offer one explanation for
discrepancies between predictions of a simple model and real
observations: if the model is a \emph{smooth} dynamical system and
predicts stable periodic motion then observed noise around the
predicted periodic motion in real observations is evidence for the
presence of (unmodelled) inhomogeneous forcing. Fenichel's Theorem
\cite{F79} guarantees that unmodelled strongly stable degrees of
freedom (that is, the fast subsystem) cannot be the cause of this
apparent noise. If the model is a Filippov system and sliding occurs
in the predicted periodic motion then the fast subsystem can disturb
the periodic motion in a way that it is no longer periodic (for
example, chaotic, or periodic with a much higher period). Hence, if
the model predicts sliding motion then apparent noise in the
observation is not evidence for unmodelled inhomogeneous forcing.

The influence of fast dynamics also provides a mechanism for chatter
(that is, rapid sequences of switches instead of sliding) which
competes with other possible mechanisms. For example, switches in
electronic circuits are known to be non-ideal \cite{BV01}. Suppose
that an electronic switch acts with a small delay $\tau$ but is
otherwise ideal. Taking this delay into account one would predict
similar effects as described in this paper
\cite{FFS02,SFF03,S06c}. Similarly, if the switches are implemented by
fast actuators (with speed $1/\tau$ as described in \cite{F02}) one
would expect to observe chattering with frequency of the order
$1/\tau$. However, if unmodelled capacities (or other stable degrees
of freedom) give rise to fast stable subsystems with decay times
$\epsilon>\tau$ then the frequency of the chatter will be limited by
$1/\epsilon$, which is smaller than $1/\tau$.

Another open question is if mechanical systems with dry friction have
a special structure that prevents singular perturbations from
destroying sliding (the case $\theta<0$ in our examples). Singular
perturbations in mechanical systems may certainly change the stability
of sliding periodic orbits near grazing. This is also possible in the
case corresponding to $\theta>0$ in our examples: sliding persists but
at grazing the one-sided derivative of the grazing orbit has a change
of order $1$ due to the fast dynamics. We demonstrate this effect in
Appendix~\ref{sec:dfo1} for the classical autonomous dry-friction
oscillator.

\section*{Acknowledgments}
Piotr Kowalczyk would like to acknowledge EPSRC grant EP/E050441/1.

\bibliographystyle{plain}
\bibliography{delay}

\begin{thebibliography}{10}

\bibitem{BV01}
S.~Banerjee and G.~Verghese.
\newblock {\em Nonlinear Phenomena in Power Electronics}.
\newblock IEEE Press, New York, 2001.

\bibitem{BKW05}
D.A.W. Barton, B.~Krauskopf, and R.E. Wilson.
\newblock Explicit periodic solutions in a model of a relay controller with
  delay and forcing.
\newblock {\em Nonlinearity}, 18(6):2637--2656, 2005.

\bibitem{DGR07}
F.~Dercole, A.~Gragnani, and S.~Rinaldi.
\newblock Bifurcation analysis of piecewise smooth ecological models.
\newblock {\em Theoretical Population Biology}, 72(2):197--213, 2007.

\bibitem{BBCK07}
M.~di~Bernardo, C.~Budd, A.R. Champneys, and P.~Kowalczyk.
\newblock {\em Piecewise-smooth Dynamical Systems {\rm ---} Theory and
  Applications}, volume 163 of {\em Applied Mathematical Sciences}.
\newblock Springer-Verlag, 2007.

\bibitem{DB09}
P.S. Dutta and S.~Banerjee.
\newblock Period increment cascades in a discontinuous map with square-root
  singularity.
\newblock In {\em Proceedings of the 2nd IFAC Conference on analysis and
  control of chaotic systems, London, U.K.}, 2009.

\bibitem{F79}
N.~Fenichel.
\newblock Geometric singular perturbation theory for ordinary differential
  equations.
\newblock {\em Journal of Differential Equations}, 31:53--98, 1979.

\bibitem{FFS02}
L.~Fridman, E.~Fridman, and E.~Shustin.
\newblock Steady modes and sliding modes in relay control systems with delay.
\newblock In J.~P. Barbot and W.~Perruquetti, editors, {\em Sliding Mode
  Control in Engineering}, pages 263--293. Marcel Dekker, New York, 2002.

\bibitem{F00}
L.M. Fridman.
\newblock Periodic motions in {VSS} and singular perturbations.
\newblock In X.~Yu and J.-X. Xu, editors, {\em Advances in Variable Structure
  Systems}, pages 365--374. World Scientific, Singapore, 2000.

\bibitem{F02}
L.M. Fridman.
\newblock Singularly perturbed analysis of chattering in relay control systems.
\newblock {\em IEEE Trans. on Automatic Control}, 47(12):2079--2084, 2002.

\bibitem{F02a}
L.M. Fridman.
\newblock Slow periodic motions with internal sliding modes in variable
  structure systems.
\newblock {\em Int. J. of Control}, 75(7):524--537, 2002.

\bibitem{H00}
R.~Horv{\'a}th.
\newblock Experimental investigation of excited and self-excited vibration.
\newblock Master's thesis, University of Technology and Economics, Budapest,
  {\url{http://www.auburn.edu/~horvaro/index2.html}}, 2000.

\bibitem{SFF03}
E.~Shustin, E.~Fridman, and L.~Fridman.
\newblock Oscillations in a second-order discontinuous system with delay.
\newblock {\em Disc. and Cont. Dyn. Sys.}, 9(2):339--358, 2003.

\bibitem{S06c}
R.~Szalai.
\newblock {\em Nonlinear dynamics of high-speed milling}.
\newblock PhD thesis, Budapest University of Technology and Economics, 2006.

\bibitem{VB99}
L.~N. Virgin and C.~J. Begely.
\newblock Grazing bifurcations and basins of attraction in an impact-friction
  oscillator.
\newblock {\em Physica D}, 130:43--57, 1999.

\end{thebibliography}

\appendix

\section{Proof of Lemma~\ref{thm:gen}}
\label{sec:lemmaproof}
The assumptions of Lemma~\ref{thm:gen} imply that the system
\begin{align}
  \dot x&=\
  \begin{cases}
    f_-(x ,y,\mu,\epsilon) & \mathrm{if\ } h(x,y,\mu,\epsilon)<0\mbox{,}\\
    f_+(x,y,\mu,\epsilon) & \mathrm{if\ } h(x,y,\mu,\epsilon)\geq0
  \end{cases}\label{eq:genslow}\\
  \epsilon\dot y &=\ g(x,y,\mu,\epsilon)\label{eq:genfast}
\end{align}
has a family of periodic orbits $(x(t;\mu,\epsilon),y(t;\mu,\epsilon))$
that grazes the manifold $\mathcal{H}_s=\{h=0\}$ for all sufficiently
small $\epsilon>0$: this follows from the fact that the reduced vector
field ($\epsilon=0$) has a family of periodic orbits
$(x(t;\mu,0),y(t;\mu,0))$ which grazes at $\mu=0$, and that the
slow-fast system \eqref{eq:genslow} and \eqref{eq:genfast} on its slow
invariant manifold $\mathcal{M}_+$ in the domain
$\mathcal{H}_+=\{h\geq0\}$ is a regular perturbation of the reduced
vector field. Without loss of generality we can (by reparametrizing
$\mu$) assume that the grazing occurs at $\mu=0$ for all $\epsilon$,
that the time at which grazing occurs on the orbit is $t=0$ for all
$\epsilon>0$, and that the point $(x(0,0,\epsilon),y(0,0,\epsilon))$ in
which the orbit grazes is $(x,y)=(0,0)$ (by shifting time of the
autonomous system and shifting the origin of the coordinate
system). In our new coordinates the family
$(x(t;\mu,\epsilon),y(t;\mu,\epsilon))$ is a family of periodic orbits,
which depends smoothly on both parameters $(\mu,\epsilon)$ in the
parameter interval
$[-\mu_\mathrm{max},0]\times[0,\epsilon_\mathrm{max}]$ (both,
$\mu_\mathrm{max}$, and $\epsilon_\mathrm{max}$ are chosen
sufficiently small). The periodic orbits are entirely inside
$\mathcal{H}_+$ and graze $\mathcal{H}_s$ quadratically in $(0,0)$ for
$\mu=0$ (for all $\epsilon\in[0,\epsilon_\mathrm{max}]$).

This quadratic grazing condition means that the function
\begin{displaymath}
\chi:(t,\mu,\epsilon)\mapsto
h(x(t;\mu,\epsilon),y(t;\mu,\epsilon),\mu,\epsilon)
\end{displaymath}
(which is smooth on
$\R\times[-\mu_\mathrm{max},0]\times[0,\epsilon_\mathrm{max}]$)
satisfies for all $\epsilon\in[0,\epsilon_\mathrm{max}]$:
\begin{enumerate}
\item\label{app:sw} $\chi(0,0,\epsilon)=0$ ($\mathcal{H}_s$ contains
  the origin for $\mu=0$)
\item\label{app:graz} $\partial_1\chi(0,0,\epsilon)=0$ (the periodic
  orbit grazes $\mathcal{H}_s$ in the origin for $\mu=0$)
\item\label{app:quad} $\partial_1^2\chi(0,0,\epsilon)>0$ (the grazing
  is quadratic)
\item\label{app:unfold} $\partial_2\chi(0,0,\epsilon)<0$ ($\mu$
  unfolds the grazing).
\end{enumerate}
In order to find a local return map of the periodic orbit
$(x(t,0,\epsilon),y(t,0,\epsilon))$ we have to choose an appropriate
Poincar{\'e} cross section: the conditions \ref{app:graz} and
\ref{app:quad} guarantee that the set
\begin{displaymath}
  \Sigma:=\{(x,y):
  \epsilon\partial_1hf_++\partial_2h\,g=0\mbox{,\ }\|(x,y)\|<\delta\}
\end{displaymath}
defines a smooth hypersurface near the origin, which intersects the
periodic orbit and the switching manifold $\mathcal{H}_s$ transversally
for all sufficiently small $\epsilon$ and $\mu$. The set $\Sigma$ is
the set of local minima of $h$ along trajectories of the flow $E_+$
generated by the vector field $(f_+,g/\epsilon)$. We define our Poincar{\'e}
section as $\Sigma_1:=E_+^{-T(\mu,\epsilon)/2}\Sigma$ where
$T(\mu,\epsilon)$ is the period of the periodic orbit and $E_+^t$ is
the time-$t$ map of the flow $E_+$.

Theorem 8.1 in textbook \cite{BBCK07} states that, if the vector field
$(f_-,g/\epsilon)$ points toward $\mathcal{H}_s(0,\epsilon)$ in the origin then
the local return map to $\Sigma_1$ is continuous and it can be
written in the form
\begin{displaymath}
  P=  E_+^{\Sigma\mapsto\Sigma_1}
\circ P_\mathrm{DM}\circ E_+^{T(\mu,\epsilon)/2}
\end{displaymath}
where $E_+^{\Sigma\mapsto\Sigma_1}$ is the map obtained by
following $E_+$ from $\Sigma$ to its first intersection with
$\Sigma_1$ (this is locally a diffeomorphism, the traveling time
is close to $T(\mu,\epsilon)/2$). The map $P_\mathrm{DM}$ maps
$\Sigma$ back to itself and accounts for all effects due to the
discontinuity (it is called the Poincar{\'e} discontinuity map in
\cite{BBCK07}). Whether the vector field $(f_-,g/\epsilon)$ points toward or
away from the switching manifold in the origin is determined by the
sign of the function
\begin{equation}\label{eq:eminus}
  \sigma_-:(x,y,\mu,\epsilon)\mapsto
  [\partial_1h\,f_-+\epsilon^{-1}\partial_2h\,g](x,y,\mu,\epsilon)
\end{equation}
in the point $(x,y,\mu,\epsilon)=(0,0,0,\epsilon)$. Due to condition
(\ref{app:graz}) we know that the time derivative of $h$ when
following $E_+$ vanishes in $(0,0,0,\epsilon)$:
\begin{equation}\label{eq:app:eplus}
\partial_1h\,f_++\epsilon^{-1}\partial_2h\,g=0\mbox{.}
\end{equation}
Thus, combining \eqref{eq:eminus} and \eqref{eq:app:eplus} we obtain
that the quantity distinguishing the two cases in Lemma~\ref{thm:gen}
is in fact
\begin{equation}
  \label{eq:chiminus}
  \sigma_-(0,0,0,\epsilon)=\partial_1h\,f_--\partial_1h\,f_+\mbox{.}
\end{equation}
That is, in case~1 the vector field $E_-$ points toward the switching
manifold $\mathcal{H}_s$ and we can apply Theorem~8.1 from
\cite{BBCK07}, proving the claim of Lemma~\ref{thm:gen} for case~1,
whereas, in case~2, $E_-$ points away from the switching manifold
$\mathcal{H}_s$. Theorem~8.1 from \cite{BBCK07} also gives an
asymptotic expression for the one-sided derivatives of the map
$P_\mathrm{DM}$ in $(0,0)$. The example from Section~\ref{sec:study} will
show that the limit of these one-sided derivatives for $\epsilon\to0$
can be different from the one-sided derivative when one applies
Theorem~8.1 from \cite{BBCK07} to the reduced system ($\epsilon=0$).

It remains to be shown that in case~2, when the vector field
$(f_-,g/\epsilon)$ points away from $\mathcal{H}_s$ in the origin for
$\mu=0$, the return map to $\Sigma_1$ is discontinuous for
$\epsilon>0$. Let us define the sets
\begin{align*}
H_1^+ &=\ E_+^{-T(\mu,\epsilon)/2} \{(x,y)\in\Sigma:h(x,y,\epsilon)\geq\epsilon\} \\
N_{\phantom{+}} &=\ E_+^{\Sigma\mapsto\Sigma_1} \{(x,y)\in
\Sigma:h(x,y,\epsilon)<0\}.
\end{align*}
%\begin{align*}
%  H_1^+&=\ (E_+^{-T(\mu,\epsilon)/2)}\{h\geq0\})\cap\Sigma_1  \\
%  N&=\ \left(\left[E_+^{\Sigma\mapsto\Sigma_1}\right]^{-1}
%    \{h<0\}\right)
%  \cap\Sigma_1\mbox{.}
%\end{align*}
The set $H_1^+$ consists of all points in the Poincar{\'e} section
$\Sigma_1$ for which $P_\mathrm{DM}$ is the identity (that is,
trajectories starting in $H_1^+$ do not switch to the flow $E_-$
during the evaluation of the return map $P$). The set $N$ consists of
points in $\Sigma_1$ that do not have a pre-image (they are images of
a repelling sliding section under $E_+^{\Sigma\mapsto\Sigma_1}$).

The set $PH_1^+\subset\Sigma_1$ can be written as
\begin{displaymath}
  PH_1^+=E_+^{\Sigma\mapsto\Sigma_1}\{(x,y)\in\Sigma:h(x,y,0)\geq0\}
\end{displaymath}
since $E_+^{\Sigma\mapsto\Sigma_1}\circ E_+^{T(\mu,\epsilon)/2}$ is
the whole return map $P$ when $P_\mathrm{DM}$ is the identity.  Thus, the set
$PH_1^+\cup N$ is an image of the whole cross-section $\Sigma$ under
the local diffeomorphism $E_+^{\Sigma\mapsto\Sigma_1}$.  Hence,
$PH_1^+\cup N$ contains an open neighborhood of the fixed point
$p_0\in\Sigma_1$ corresponding to the grazing periodic
orbit. Moreover, $P$ is a diffeomorphism between $H_1^+$ and
$PH_1^+$. This implies that points from $\Sigma_1\setminus H_1^+$
cannot be mapped into the neighborhood of $p_0$ under $P$ (they can
neither be mapped into $N$ nor into $PH_1^+$). However, $p_0$ is on
the boundary of $H_1^+$ and, thus, it is in the closure of
$\Sigma_1\setminus H_1^+$. Consequently, there exist points
arbitrarily close to $p_0$ that do not get mapped into the
neighborhood of $p_0$ under $P$, which proves that $P$ is
discontinuous in case~2.

We note that the proof of discontinuity in case~2 of
Lemma~\ref{thm:gen} is not specific to slow-fast systems (we only use
that the vector field $E_-$ points away from the switching manifold at
the grazing point $(0,0,0,\epsilon)$) but it relies on the fact that
the grazing orbit lies entirely in the domain governed by the flow
$E_+$ (that is why $P$ restricted to $H_1^+$ is a diffeomorphism). In
general, for example, if the grazing periodic orbit has sliding
segments of positive length in other parts of the phase space, the
return map may not be discontinuous in case~2.

\section{The Poincar{\'e} discontinuity mappings of the minimal
  example in Section~\ref{sec:study}}

This appendix derives the dominant terms of the projected
discontinuity mapping
\begin{displaymath}
P_\mathrm{DM}=P_3\circ P_2
\end{displaymath}
as defined for the minimal example in Section~\ref{sec:study}.  The
map $P_\mathrm{DM}$ maps the line $\mathcal{
  L}_2=\{h_+'(x,y)=0\mbox{,\ } x_1=y\}$ back to itself in the
neighborhood of the grazing point $p_g=(-1,0,-1)$.  Orbits of $E_+$
crossing $\mathcal{ L}_2$ in a point with coordinate $x_1\geq-1$ do
not switch to $E_-$ or slide: by definition of $\mathcal{ L}_2$ each
orbit has its local minimum of $h$ exactly when it crosses $\mathcal{
  L}_2$, and $h$ is identical to $x_1+1$ on the invariant manifold
$\mathcal{ M}_+$ of $E_+$.  Thus, $\mathcal{L}_2$ can be parametrized
by the coordinate $x_1$, and for $x_1\geq-1$ the map $P_\mathrm{DM}$
is the identity: $P_\mathrm{DM}( x_1)= x_1$. For $x_1<-1$ the map
$P_\mathrm{DM}$ is itself again a composition of several maps:
\begin{equation}\label{eq:pdmdiscdef}
  P_\mathrm{DM}=\Pi_3\circ P_3\circ\Pi_2\circ\Pi_1
\end{equation}
where
\begin{itemize}
\item $\Pi_1$ follows the flow $E_+$ backward in time from $\mathcal{
    L}_2$ to the switching manifold $\mathcal{ H}_s$, specifically, the
  line $\mathcal{ L}_3=\{\tilde x_1=\tilde y=0\}$,
\item the map $\Pi_2$ depends on the sign of the parameter $\theta$:
  \begin{description}
  \item[$\theta<0$:] $\Pi_2$ follows $E_-$ from $\mathcal{ L}_3$
    forward in time until one hits the switching manifold $\mathcal{
      H}_s$ again;
  \item[$\theta>0$:] $\Pi_2$ follows the sliding flow $E_s$, defined
    by \eqref{eq:slide}, starting from $\mathcal{ L}_3$ and staying in
    the switching surface $\mathcal{ H}_s$ until the flow $E_+$
    becomes tangent to $\mathcal{ H}_s$, that is, $h_+'(x,y)$ becomes
    zero;
  \end{description}
\item $P_3$ projects along the stable fibre of $E_+$ down onto the
  slow invariant manifold $\mathcal{ M}_+$:
  $P_3(x_1,x_2,y)=(x_1,x_2,x_1)$.
\item $\Pi_3$ follows $E_+$ on $\mathcal{ M}_+$ (forward or backward in
  time) to $\mathcal{ L}_2$.
\end{itemize}
We have swapped $P_3$ and $\Pi_3$ in \eqref{eq:pdmdiscdef}, which is
possible because $P_3$ and $\Pi_3$ commute.

\subsection{The case  $\theta<0$}
\label{sec:discont}
\begin{figure}[t]
  \centering
  \includegraphics[scale=1]{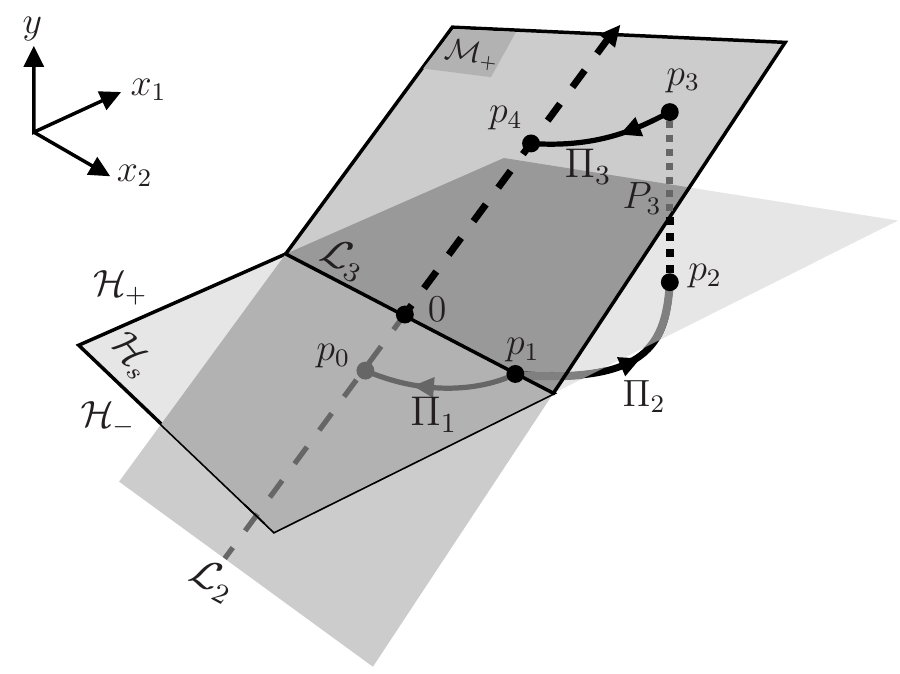}
  \caption{The Poincar{\'e} discontinuity map for the case $\theta<0$
    near the point $p_g$ ($0$ in the rescaled coordindate system
    \eqref{eq:scal}).}
  \label{fig:pdmthlt0}
\end{figure}
For $\theta<0$ the parameter range of interest is of order $\epsilon$.
Similarly, the scale of the dynamics transversal to the limit cycle
(that is, the $x_1$ and the $y$ components) is of order $\epsilon$. We
zoom into the neighborhood of the grazing point $p_g=(-1,0,-1)$ by
introducing the new quantities $\xi$, $\psi$ and $q$ (which are
rescalings of the coordinates $x$, $y$ and
$\mu$):
\begin{equation}\label{eq:scal}
  \begin{aligned}
    \epsilon&=\ \phantom{-}\delta^2&
    x_1&=\ -1+\delta^2\xi_1\\
    \mu&=\ \phantom{-}1+\delta^2 q\qquad&
    x_2&=\ \phantom{-}\delta\xi_2\\
    &&y&=\ -1+\delta^2\psi\mbox{.}\\
  \end{aligned}
\end{equation}
Any point in the phase-parameter space with $\xi$, $\psi$ and $q$ of
order $1$ is in the vicinity of the grazing bifurcation. The scaling
\eqref{eq:scal} squeezes the direction along the grazing periodic
orbit by scaling it only by $\delta=\sqrt{\epsilon}$. In the new
coordinates the switching manifold $\mathcal{ H}_s$ is $\{(\xi,\psi):
\theta\xi_1+(1-\theta)\psi=0\}$.
System~\eqref{eq:exslow},\,\eqref{eq:exfast} in the new coordinates
has the form (with rescaled time $t_{\mathrm{new}}=
t_{\mathrm{old}}/\delta$ and truncating terms of order $\delta^3$):
\begin{align}
  \dot\xi&=
  \begin{cases}
    \begin{bmatrix}
      -\omega\xi_2+[\xi_2^2-2\xi_1-q]\delta+O(\delta^3)\\
      -\omega+\omega\delta^2\xi_1+O(\delta^3)
    \end{bmatrix}%\hfill\qquad\  \\
    %\hfill
    &\mbox{if $\theta\xi_1+(1-\theta)\psi\geq0$}\\
    \begin{bmatrix}
      1/\delta\\ 0
    \end{bmatrix}&\mbox{if $\theta\xi_1+(1-\theta)\psi<0$}\\
  \end{cases}\label{eq:xidot}\\
  \delta\dot\psi&=
  -\psi+\xi_1-\omega\xi_2\delta+[\xi_2^2-2\xi_1-q]\delta^2
  +O(\delta^4)\mbox{.}\label{eq:psidot}
\end{align}
We expand the quantities $h_+'$ and $h_-'$ in the new coordinates in
$\delta$ (dropping $O(\delta^3)$ and substituting $\psi$ by
$\theta\xi_1/(\theta-1)$):
\begin{equation}\label{eq:tangency}
  \begin{split}
    \delta^{-1} h_+'&=\ \xi_1-\omega\xi_2\delta+
    \left[\xi_2^2-q-2\xi_1\right]\delta^2+O(\delta^3)\\
    \delta^{-1} h_-'&=\ \theta+\xi_1+(\theta-1)\omega\xi_2\delta\\
    &\phantom{=\ } +(1-\theta)
    \left[\xi_2^2-q-2\xi_1\right]\delta^2+O(\delta^3)\mbox{.}
  \end{split}
\end{equation}
Thus, in the vicinity of the grazing point the vector field $E_+$
points toward the switching manifold $\mathcal{ H}_s$ if $h_+'<0$, that is,
\begin{equation}
  \label{eq:epslides}
  \xi_1< \omega \xi_2\delta-\left[\xi_2^2-q\right]\delta^2
  +O(\delta^3)\mbox{,}
\end{equation}
and $E_-$ points toward $\mathcal{ H}_s$ if $h_-'>0$, that is,
\begin{equation}
  \label{eq:emslides}
  \xi_1> -\theta+(1-\theta)\omega\xi_2\delta +(1-\theta)
    \left[q-\xi_2^2-2\theta\right]\delta^2+O(\delta^3)\mbox{.}
\end{equation}

The curve $\mathcal{ L}_2$ is nearly linear for $\xi$ of order $1$. It can
be parametrized in the rescaled coordinates $[\xi_1,\xi_2,\psi]$ (see
\eqref{eq:scal}) by the map
\begin{equation}
  \label{eq:l2par}
  L_2: \R\mapsto\R^3: \xi_1\mapsto
    [\xi_1,
    \eta(\xi_1),
    \xi_1]
\end{equation}
where the second component ($\xi_2$) of $L_2$ is the locally unique
solution $\xi_2$ of the equation
\begin{equation}\label{eq:l2def}
    h_+'(\xi_1,\xi_2,\xi_1,q,\delta)=0
\end{equation}
depending on $\xi_1$, $q$ and $\delta$ (the grazing is quadratic,
hence $h_+'$ is monotone in $\xi_1$). More precisely, the solution
$\xi_2$ of \eqref{eq:l2def} can be represented as a graph
$\eta(\xi_1)$ (dropping the dependence of $\eta$ on $q$ and $\delta$),
and this graph can be expanded in $\delta$:
\begin{equation}
  \label{eq:etaseries}
  \eta(\xi_1)=-\frac{2\xi_1+q}{\omega}\delta+
  \delta^3r(\xi_1,q,\delta)[\xi_1,q]\mbox{.}
\end{equation}
The remainder term of order $\delta^3$ in \eqref{eq:etaseries} is at
most linear in $\xi_1$ and $q$ because $\eta(0)=0$ for $\xi_1=q=0$ and all
$\delta$ (the grazing point $\xi_1=\xi_2=\psi=q=0$ satisfies
\eqref{eq:l2def} for all $\delta$).

\paragraph*{The map $\Pi_1$} We expand the trajectory
$(\xi_1(t),\xi_2(t),\psi(t))$ of $E_+$ through a point
\begin{displaymath}
p_0=\left(\xi_{1,0},\eta(\xi_{1,0}),\xi_{1,0}\right)\in\mathcal{ L}_2
\end{displaymath}
in time ($\psi(t)=\xi_1(t)$ because $p_0\in\mathcal{ M}_+$):
\begin{displaymath}
%  \label{eq:p1texp}
  \begin{split}
    \xi(t)=\ &
    \begin{bmatrix}
      \xi_{1,0}\\ -\frac{2\xi_{1,0}+q}{\omega}\delta+O(\delta^3)
    \end{bmatrix}
    +
    \begin{bmatrix}
      0\\-\omega+O(\delta^2)
    \end{bmatrix}t+
    \begin{bmatrix}
      \frac{\omega^2}{2}+O(\delta^2)\\ O(\delta^3)
    \end{bmatrix}t^2+t^3O(\delta^2)\mbox{.}
  \end{split}
\end{displaymath}
The coefficient in front of $t^2$ is non-zero in its first component
because we have quadratic grazing. The first component of the
coefficient in front of $t$ vanishes by choice of $\mathcal{ L}_2$.
Thus, for $\xi_{1,0}<0$ and $|\xi_{1,0}|$ of order $1$ or less, the
trajectory intersects the switching manifold $\{\xi_1=0\}$ at
\begin{equation}
  \label{eq:s1int}
  t=\pm\frac{\sqrt{-2\xi_{1,0}}}{\omega}\,(1+O(\delta^2))+
  O(\delta^2)\xi_{1,0}+O(\delta^3)\mbox{.}
\end{equation}
The map $\Pi_1$ is defined by following $E_+$ backward in time. Thus,
we choose the negative sign in \eqref{eq:s1int}. This gives rise to an
intersection of the trajectory with the switching surface at a point
$(0,\xi_{2,1},0)$ with the coordinate
\begin{equation}
  \label{eq:xi2int}
  \xi_{2,1}=\sqrt{-2\xi_{1,0}}\,(1+O(\delta^2))-\frac{2\xi_{1,0}+q}{\omega}\delta
  +O(\delta^2)\mbox{.}
\end{equation}
The first term of the sum in \eqref{eq:xi2int} contains all
square-root terms of $\xi_{1,0}$. Consequently the
map $\Pi_1$ maps
\begin{displaymath}
  \Pi_1: p_0=[\xi_{1,0},\eta(\xi_{1,0}),\xi_{1,0}]\!\in\!\mathcal{ L}_2
  \mapsto p_1=[0,\xi_{2,1},0]\!\in\!\mathcal{ L}_3
\end{displaymath}
where $\xi_{2,1}$ is given by \eqref{eq:xi2int}. The dominant term in
the expansion of $p_1$ with respect to the coordinate $\xi_{1,0}$ of
$p_0$ is a square root.

\paragraph*{The map $\Pi_2$} For $\xi_{1,0}$ of order $1$ (and, thus,
$\xi_{2,1}$ of order $1$) the flow $E_-$ points away from the
switching surface $\mathcal{ H}_s$ in $p_1$ because condition
\eqref{eq:emslides} is not satisfied in $p_1$ (because
$\theta<0$). The initial point for the map $\Pi_2$ is
$p_1=(0,\xi_{2,1},0)$. The trajectory $(\xi_1(s),\xi_2(s),\psi(s))$
following $E_-$ through this point satisfies
\begin{equation}\label{eq:fm:rescaled}
  \begin{split}
    \xi_1(s)&=\ s\\
    \xi_2(s)&=\ \xi_{2,1}\\
    \psi(s)&=\ s+\exp(-s)-1+\omega\xi_{2,1}[\exp(-s)-1]\delta+O(\delta^2)
  \end{split}
\end{equation}
where $s=t/\delta$ is the rescaled time. There is a unique
time $s_*(\xi_{2,1})>0$ such that the trajectory hits the switching
surface $\{\theta\xi_1+(1-\theta)\psi=0\}$ again. The expansion of
$s_*(\xi_{2,1})$ in $\delta$ is
\begin{align}
  s_*(\xi_{2,1})&=\ s_0(\theta)+\delta\,\frac{\omega
    s_0(\theta)}{\theta+s_0(\theta)}\,\xi_{2,1}+O(\delta^2)\label{eq:hittime}
\end{align}
and $s_0(\theta)$ is the unique positive solution of the equation
\begin{equation}
  \label{eq:s0def}
  (1-\theta)\exp(-s_0)-1+s_0+\theta=0\mbox{.}
\end{equation}
\begin{figure}[t]
  \centering
  \includegraphics[width=0.25\textwidth]{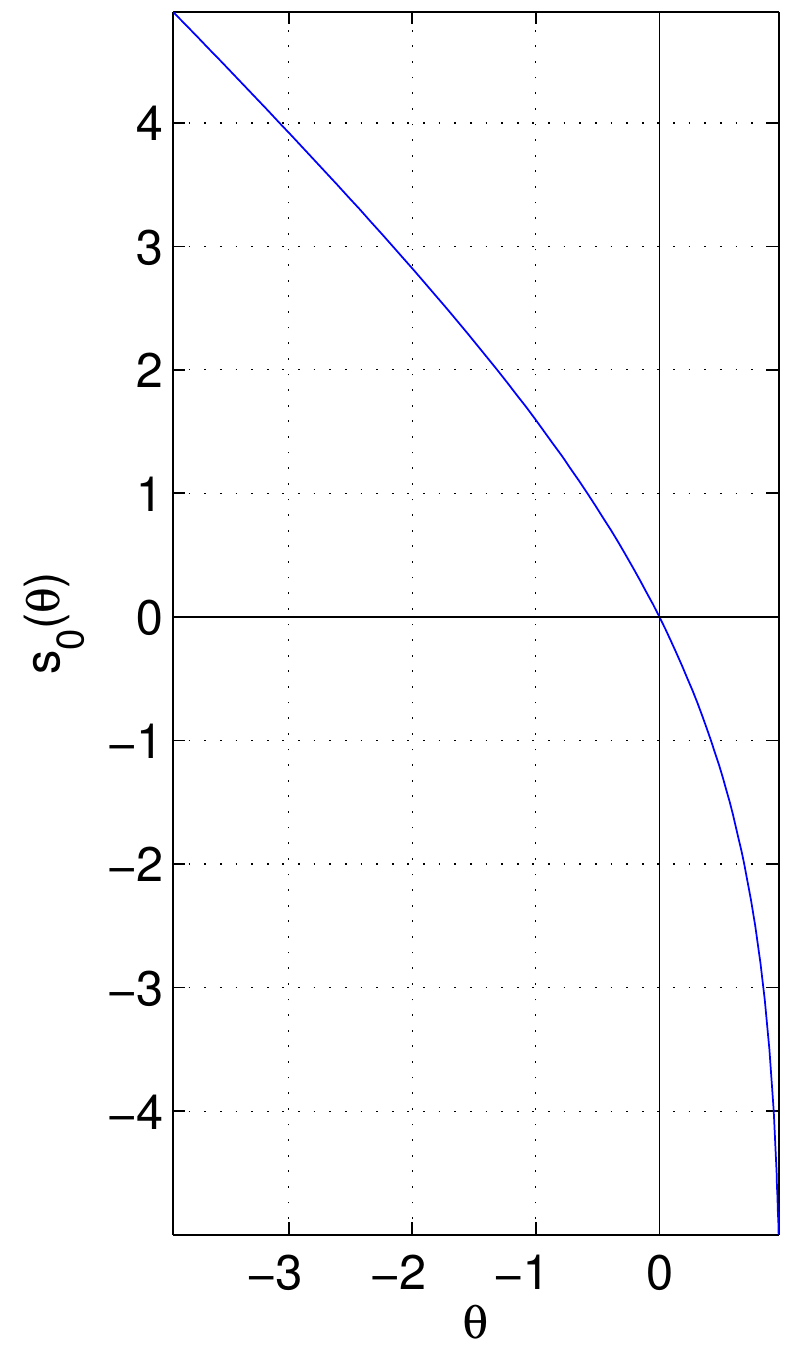}
  \caption{Implicitly given graph of $s_0(\theta)$.}
  \label{fig:s0theta}
\end{figure}
Figure~\ref{fig:s0theta} shows the graph of $s_0(\theta)$, which can
be determined implicitly by expressing $\theta$ as a function of
$s_0$. Since we assume that $|\theta|$ is of order $1$ we cannot
replace \eqref{eq:hittime} by its expansion in $\theta$. The
denominator in the coefficient for $\delta$ in \eqref{eq:hittime} is
positive because the solution $s_0$ of \eqref{eq:s0def} is greater than
$-\theta$ for all $\theta<0$.  Consequently, $\Pi_2$ maps
\begin{displaymath}
  \Pi_2: p_1=(0,\xi_{2,1},0)\mapsto
  p_2=\left[s_*(\xi_{2,1}),\xi_{2,1},
    \frac{-\theta s_*(\xi_{2,1})}{1-\theta}\right]\mbox{.}
\end{displaymath}
For small $\delta$, and $|\xi_{1,2}|$ of order $1$ or less, the sliding
condition \eqref{eq:epslides} is not satisfied for points in the image
of $\Pi_2$ because $s_0(\theta)>0$ is of order $1$. This implies that
the flow will switch to $E_+$.  The projection $P_3$ of the image
along the stable fibres of $E_+$
\begin{displaymath}
  P_3: p_2=\left[s_*(\xi_{2,1}),\xi_{2,1},\frac{-\theta
      s_*(\xi_{2,1})}{1-\theta}\right]
  \mapsto
  p_3=\left[s_*(\xi_{2,1}),\xi_{2,1},s_*(\xi_{2,1})\right]
\end{displaymath}
maps into the slow invariant manifold $\mathcal{ M}_+$ of $E_+$ and
away from the switching manifold because $0<-\theta/(1-\theta)<1$ and
$s_*(\xi_{2,1})>0$. Thus, the point $p_3$ is also in the half space
$\mathcal{ H}_+=\{(\xi,\psi): \theta\xi_1+(1-\theta)\psi>0\}$.

\paragraph*{The map $\Pi_3$} The map $\Pi_3$ maps the point $p_3$ of
the slow invariant manifold $\mathcal{ M}_+$ back to the Poincar{\'e}
section (more precisely its intersection with $\mathcal{ M}_+$), $\mathcal{
  L}_2$, by following $E_+$. This can be obtained in the same manner
as the map $\Pi_1$: denote the components of $p_3$ by
$[\xi_{1,3},\xi_{2,3},\psi_3]$ (where $\psi_3=\xi_{1,3}$ because
$p_3\in\mathcal{ M}_+$), and denote $\xi$-component of the trajectory
$E_+^tp_3$ by $(\xi_1(t),\xi_2(t))$.  The time $t_3$ when the
trajectory hits $\mathcal{ L}_2$ has to satisfy the condition defining
$\mathcal{ L}_2$
\begin{equation}
\eta(\xi_1(t_3))=\xi_2(t_3)\mbox{,}\label{eq:l2hit}
\end{equation}
which defines $t_3$ and the point
$p_4=(\xi_{1,4},\xi_{2,4},\psi_4)=(\xi_1(t_3),\xi_2(t_3),\xi_1(t_3))$
implicitly. We expand $\xi_1(t)$ and $\xi_2(t)$ with respect to time
$t$ in $0$ to second order in $t$ using \eqref{eq:xidot} (all
third-order terms have a coefficient of order $\delta^3$) and
\eqref{eq:etaseries}, and insert the expansion of $t_3$ in $\delta$,
\begin{equation}\label{eq:t3series}
  t_3=t_{3,0}+\delta t_{3,1}+O(\delta^2)\mbox{,}
\end{equation}
into \eqref{eq:l2hit} where $t_{3,0}$ and $t_{3,1}$ are the unknowns
at zero and first order level. Solving \eqref{eq:l2hit} to first order in $\delta$ gives
\begin{align}
    t_3&=\
    \frac{\xi_{2,3}}{\omega}+\frac{q+2\xi_{1,3}}{\omega^2}\delta+O(\delta^2)\mbox{\quad
      and, thus,}\nonumber\\
    \xi_{1,4}&=\ \xi_{1,3}+\frac{\xi_{2,3}^3}{\omega}\delta+O(\delta^2)\mbox{.}\label{eq:xi14}
\end{align}
Consequently, the map $\Pi_3$ maps
\begin{equation}\label{eq:pi3}
  \Pi_3:p_3=[\xi_{1,3},\xi_{2,3},\xi_{1,3}]\mapsto
  p_4=[\xi_{1,4},\eta(\xi_{1,4}),\xi_{1,4}]\!\in\mathcal{ L}_2
\end{equation}
where $\xi_{1,4}$ is given \eqref{eq:xi14}.

\paragraph*{The PDM}
Combining the expressions \eqref{eq:xi2int}, \eqref{eq:hittime} and
\eqref{eq:xi14} we obtain the composition $P_{\mathrm{DM}}=\Pi_3\circ P_3\circ
\Pi_2\circ\Pi_1$:
\begin{equation}
  \label{eq:pdmdisc}
  P_{\mathrm{DM}}(\xi_1)=
  \begin{cases}
    \xi_1 &\mbox{if $\xi_1\geq0$}\\
    s_0(\theta)+\delta\left[\frac{\omega
    s_0(\theta)}{\theta+s_0(\theta)}\sqrt{-2\xi_1}+
      \frac{(-2\xi_1)^{3/2}}{\omega}\right]+O(\delta^2)
    & \mbox{if $\xi_1<0$.}
  \end{cases}
\end{equation}
We have expressed the map $P_\mathrm{DM}$ as a map for the coordinate
$\xi_1$ of the point on the curve $\mathcal{ L}_2$. Given the
coordinate $\xi_1$, the other coordinates of a point of $\mathcal{
  L}_2$ are $\xi_2=\eta(\xi_1)$ and $\psi=\xi_1$ where the expansion
of $\eta$ is given in \eqref{eq:etaseries}. The expression for
$s_0(\theta)$ is given implicitly by \eqref{eq:s0def}. The
remainder term $O(\delta^2)$ for $\xi_1<0$ may contain small
corrections to the constant and square-root terms in $\xi_1$. However,
the form \eqref{eq:pdmdisc} guarantees that for sufficiently small
$\delta$ (and, thus, singular perturbation parameter $\epsilon$) the
overall return map $P_g\circ P_\mathrm{DM}$ changes from a piecewise
linear map for the reduced model ($\epsilon=0$) to a discontinuous map
for the full model ($\epsilon>0$). The size of the jump at the
discontinuity is of order $\epsilon$ in the original
coordinates. Moreover, the slope of the map next to the discontinuity
is infinity from one side.

\subsection{The case $\theta>0$}
\label{sec:slid}
\begin{figure}[t]
  \centering
  \includegraphics[scale=1]{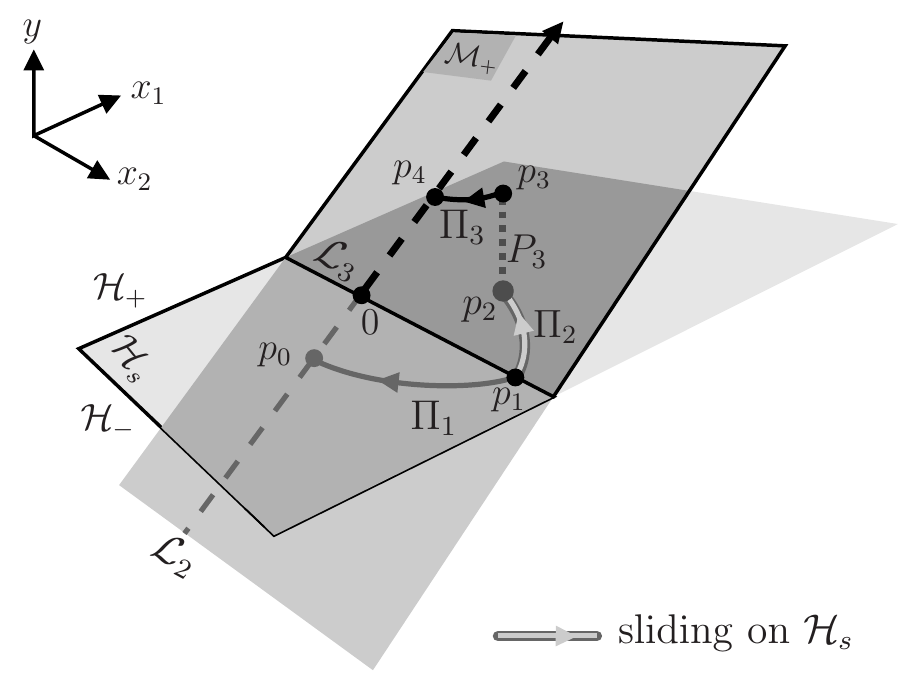}
  \caption{The Poincar{\'e} discontinuity map for the case $\theta>0$
    near the point $p_g$ ($0$ in the rescaled coordindate system
    \eqref{eq:scal}).}
  \label{fig:pdmthgt0}
\end{figure}
For $\theta>0$ the assumptions behind the validity for the asymptotic
form of the PDM near a grazing-sliding orbit are satisfied for the
singularly perturbed system \cite{BBCK07}. The flow $E_+$ is grazing
the switching surface $\mathcal{H}_s=\{\theta x_1+(1-\theta) y=0\}$
quadratically in the point $p_g=[-1,0,-1]^T$. The other flow $E_-$
points toward the switching surface near the grazing point $p_g$ of
$E_+$.  Thus, according to Theorem 8.1 of \cite{BBCK07} we have a
generic grazing-sliding bifurcation. In general, the zero-time
discontinuity map $P_\mathrm{Z}$ (ZDM) is piecewise asymptotically
linear, having the form (ignoring higher-order terms in $v$)
\begin{equation}
  \label{eq:p:zdmgen}
  P_\mathrm{Z}(p_g+v)=p_g+
  \begin{cases}
    v & \mbox{if $\partial h\, v\geq0$,}\\
    v-\frac{F_--F_+}{\partial h\,F_-}\,\partial h\,v & \mbox{if $\partial h\,v<0$}
  \end{cases}
\end{equation}
where $v\in\R^3$ is small, $p_g$ is the grazing point,
$\partial h$ is the normal of the switching surface, and
$F_\pm=(f_\pm,g/\epsilon)$ is the derivative of $E_\pm$ in $p_g$. In
our example we have $p_g=[-1,0,-1]^T$, $\partial
h=[\theta,0,1-\theta]$, $F_+=[0,-\omega,0]^T$ and $F_-=[1,0,0]^T$ at
parameter $\mu=1$.  Thus, at $\mu=1$ the linear approximation of the ZDM,
$P_\mathrm{Z}$, for points in $\mathcal{ M}_+$, the invariant manifold
of $E_+$, is
\begin{equation}
  \label{eq:p:zdm}
  P_\mathrm{Z}
  \begin{bmatrix}
    x_1\\ x_2\\ x_1
  \end{bmatrix}
  =
%   \begin{bmatrix}
%     -1\\ 0\\ -1
%   \end{bmatrix}
%   +
  \begin{cases}
  \begin{bmatrix}
    x_1\\ x_2\\ x_1
  \end{bmatrix}
     & \mbox{if $x_1\geq -1$,}\\
     \begin{bmatrix}
       (1-\theta^{-1})(x_1+1)-1\\
       x_2-\omega\theta^{-1} (x_1+1)\\
       x_1
     \end{bmatrix}
     & \mbox{if $ x_1<-1$.}
  \end{cases}
\end{equation}
The map $P_\mathrm{Z}$ can be converted to the projected PDM $P_\mathrm{DM}$ by
applying $P_\mathrm{Z}$ to points in $\mathcal{ L}_2$, then applying
the projection $P_3$ along the stable fibres of $E_+$, and finally
applying the map $\Pi_4$, which is the same as for the case
$\theta<0$, following $E_+$ on its invariant manifold $\mathcal{ M}_+$
back to $\mathcal{ L}_2$. Thus, ignoring terms of order $x_1^{3/2}$ or higher,
the PDM maps the coordinate $x_1$ of a point in $\mathcal{ L}_2$ as
follows:
\begin{equation}
  \label{eq:p:pdm}
  P_\mathrm{DM}(x_1)=
  \begin{cases}
    x_1 & \mbox{if $x_1\geq -1$}\\
    \left[1-\theta^{-1}\right]\,(x_1+1)-1 &\mbox{if $ x_1<-1$.}
  \end{cases}
\end{equation}
The slope is different for $x_1<-1$ from the slope of the reduced
model ($\epsilon=0$, where $P_\mathrm{DM}(x_1)=-1$ for $ x_1<-1$) if
$\theta\in(0,1)$.  The difference in the slope is uniform for all
small $\epsilon>0$ (expression \eqref{eq:p:pdm} does not depend on
$\epsilon$). For parameters $\mu$ close to $1$ the slope of
$P_\mathrm{DM}$ for $x_1<-1$ changes at most by a term of order $\mu-1$.
The slope of $P_\mathrm{DM}$ for $x_1<-1$ is negative for
$\theta\in(0,1)$ and can be arbitrarily large in modulus, thus, giving
rise to a strong expansion for small $\theta$ (and, possibly, chaos)
in a small neighborhood of the grazing orbit.

The remainder of this section will show that the size of this
neighborhood where the slope of $P_\mathrm{DM}$ is negative and
non-small is at least of order $\epsilon^2$ (judging from
Figure~\ref{fig:retmaps}(b) it is larger). In order to see this we
have to obtain the expressions for the maps $\Pi_1$, $\Pi_2$ and
$\Pi_3$ for this case as well. The maps $\Pi_1$ and $\Pi_3$ are
identical to the case $\theta<0$ apart from a slightly differently
scaled set of coordinates. The map $\Pi_2$ follows the sliding flow
$E_s$ to the curve $\{h_+'(x,y)=0, h(x,y)=0\}$ in the case $\theta>0$,
and is, thus, different from the map $\Pi_2$ for the case $\theta<0$.

We zoom into the neighborhood of the grazing point by introducing the
new quantities $\xi$, $\psi$ and $q$ with a scaling $\epsilon^2$
(different from the scaling for $\theta<0$):
\begin{equation}\label{eq:p:scal}
  \begin{aligned}
    x_1&=\ -1+\epsilon^2\xi_1\mbox{,}\qquad   &
    y&=\ -1+\epsilon^2\psi\mbox{,}\\
    x_2&=\ \phantom{-}\epsilon\xi_2\mbox{,}& \mu&=\
    \phantom{-}1+\epsilon^2 q\mbox{.}
  \end{aligned}
\end{equation}
Similarly as for $\theta<0$, any point in the phase-parameter space
with $\xi$, $\psi$ and $q$ of order $1$ is in the vicinity of the
grazing bifurcation. The switching manifold $\mathcal{ H}_s$ is
$\{(\xi,\psi): \theta\xi_1+(1-\theta)\psi=0\}$. System
\eqref{eq:exslow},\,\eqref{eq:exfast} in the rescaled coordinates has
the form (with rescaled time $t_{\mathrm{new}}=
t_{\mathrm{old}}/\epsilon$ and truncating terms of order
$\epsilon^3$):
\begin{align}
  \dot\xi&=
  \begin{cases}
    \begin{bmatrix}
      -\omega\xi_2+[\xi_2^2-2\xi_1-q]\epsilon+O(\epsilon^3)\\
      -\omega+\omega\epsilon^2\xi_1+O(\epsilon^3)
    \end{bmatrix}&\mbox{if $\theta\xi_1+(1-\theta)\psi\geq0$}\\
    \begin{bmatrix}
      1/\epsilon\\ 0
    \end{bmatrix}&\mbox{if $\theta\xi_1+(1-\theta)\psi<0$}\\
  \end{cases}\label{eq:p:xidot}\\
  \dot\psi&=
  -\psi+\xi_1-\omega\xi_2+[\xi_2^2-2\xi_1-q]\epsilon+O(\epsilon^3)\label{eq:p:psidot}
\end{align}
For system \eqref{eq:p:xidot},\,\eqref{eq:p:psidot} each trajectory
can spend at most a time span of order $1$ in any bounded ball (of
size of order $1$) in the $(\xi,\psi)$-coordinates. The expansions of
the quantities $h_+'$ and $h_-'$ (determining where the system is
sliding) in $\epsilon$ are
\begin{equation}\label{eq:p:tangency}
  \begin{split}
    \epsilon^{-2} h_+'&=\ \xi_1-\omega\xi_2+
    \left[\xi_2^2-q-2\xi_1\right]\epsilon+O(\epsilon^3)\\
    \epsilon^{-2} h_-'&=\
    \epsilon^{-1}\theta+\xi_1+(\theta-1)\omega\xi_2+O(\epsilon)\mbox{.}
  \end{split}
\end{equation}
As the expression for $h_-'$ is always positive for $\xi$, $\psi$ and
$q$ of order $1$ the system slides on $\mathcal{ H}_s$ whenever $h_+'<0$.

The expressions for the curve $\mathcal{ L}_2$ and the maps $\Pi_1$ and
$\Pi_3$ are identical to \eqref{eq:l2par}, \eqref{eq:etaseries},
\eqref{eq:xi2int} and \eqref{eq:xi14}, \eqref{eq:pi3}. The only
modification is the differently scaled set of coordinates
\eqref{eq:p:scal}. The effect of this scaling is that we have to
replace $\delta$ by $\epsilon$ in all equations from \eqref{eq:l2def}
to \eqref{eq:xi2int} and in \eqref{eq:xi14}, leaving all expressions
identical otherwise.  Thus, $\Pi_1$ and $\Pi_3$ map
\begin{align*}
  \Pi_1:&\ p_0=[\xi_{1,0},\eta(\xi_{1,0}),\xi_{1,0}]\in\mathcal{ L}_2
  \mapsto p_1=[0,\xi_{2,1},0]\!\in\!\mathcal{ L}_3\\
  \Pi_3:&\ p_3=[\xi_{1,3},\xi_{2,3},\xi_{1,3}] \mapsto
  p_4=[\xi_{1,4},\eta(\xi_{1,4}),\xi_{1,4}]\in\mathcal{ L}_2
\end{align*}
where $\eta$, $\xi_{2,1}$ and $\xi_{1,4}$ are given by
\begin{align*}
  \eta(\xi_1)&=\ -\frac{2\xi_1+q}{\omega}\epsilon+
  \epsilon^3r(\xi_1,q,\epsilon)[\xi_1,q]\mbox{,}\\
  \xi_{2,1}&=\ \sqrt{-2\xi_{1,0}}\,(1+O(\epsilon^2))
  -\frac{2\xi_{1,0}+q}{\omega}\epsilon
  +O(\epsilon^2)\mbox{,}\\
  \xi_{1,4}&=\
  \xi_{1,3}+\frac{\xi_{2,3}^3}{\omega}\epsilon+O(\epsilon^2)\mbox{.}
\end{align*}

\paragraph*{The map $\Pi_2$}
For $\xi_{1,0}$ (and, thus, $\xi_{2,1}$) of order $1$ the flow $E_-$
points toward the switching surface $\mathcal{ H}_s$. In $p_1$ the
flow $E_+$ points toward the switching surface, too (as $h_+'<0$ in
$p_1$), such that $\Pi_2$ is defined by following the sliding flow
$E_s^t$ until $h_+'$ (given by \eqref{eq:p:tangency}) becomes
zero. The sliding flow $E_s^t$ is invariant on the switching manifold
$\mathcal{ H}_s$. It follows an ODE for $\xi$ ($\psi$ is given by
$\psi=-\theta/(1-\theta)\xi_1$). The right-hand-side of the ODE can be
expanded in $\epsilon$ as
\begin{equation}
  \label{eq:fs}
  \begin{split}
    \dot \xi_1&=\ -\frac{1}{\theta}[\xi_1-\omega\xi_2(1-\theta)]+O(\epsilon)\\
    \dot \xi_2&=\ -\omega +O(\epsilon)\mbox{.}
  \end{split}
\end{equation}
The initial condition is $(0,\xi_{2,1})$ (where $\xi_{2,1}>0$). The
time $t$ that the system follows \eqref{eq:fs} until $h_+'$ becomes
zero is of order $1$ such that the trajectory $\xi(t)$ is a
perturbation of order $\epsilon$ of
\begin{equation}\label{eq:p:pi2flow}
  \begin{split}
    \xi_1(t)&=\
    \omega(1-\theta)\left(\left[1-\exp(-t/\theta)\right]
      (\theta\omega+\xi_{2,1})-\omega t\right)\\
    \xi_2(t)&=\ \xi_{2,1}-\omega t\mbox{.}
  \end{split}
\end{equation}
We insert the trajectory \eqref{eq:p:pi2flow} into the expansion
\eqref{eq:p:tangency} for $h_+'$ to find the unique positive time
$t_2$ for which $\epsilon^{-2}h_+'$ in \eqref{eq:p:tangency} becomes
zero in a point
\begin{displaymath}
  p_2=\left(\xi_{1,2},\xi_{2,2},\frac{\theta}{\theta-1}\,\xi_{1,2}\right)\mbox{.}
\end{displaymath}
This time $t_2$, the time for which for the system follows the sliding
flow \eqref{eq:fs}, is given implicitly to zero order in $\epsilon$ by
the equation
\begin{equation}
  \label{eq:p:xi21}
  \xi_{2,1}=
  \omega\theta\,\frac{(1-\theta)\left[1-\exp(-t_2/\theta)\right]
    +t_2}{\theta+(1-\theta)\exp(-t_2/\theta)}\mbox{.}
\end{equation}
The right-hand-side in \eqref{eq:p:xi21} is uniformly monotone
increasing in $t_2$, and it is zero for $t_2=0$. Thus,
\eqref{eq:p:xi21} can be solved for $t_2$. In combination with the
zero-order approximations
\begin{align}
  \xi_{1,2}&=\ \omega^2(1-\theta)\,
  \frac{\theta-\left[\theta+t_2\right]\exp(-t_2/\theta)}{\theta+(1-\theta)\exp(-t_2/\theta)}
    \label{eq:p:xi12}\\
    \xi_{2,2}&=\ \xi_{1,2}/\omega\label{eq:p:xi22}
\end{align}
for $\xi_{1,2}$ and $\xi_{2,2}$ one obtains well defined graphs for
$\xi_{1,2}(\xi_{2,1})$ and $\xi_{2,2}(\xi_{2,1})$. As monotonicity persists under small
perturbations these graphs remain well-defined also if we take all
orders of $\epsilon$ into account.  % Moreover, the general theory for
% grazing-sliding bifurcations states that
% \begin{displaymath}
%   \xi_{1,2}(\xi_{2,1})=\eta(0)+O((\xi_{2,1}-\eta(0))^2)
% \end{displaymath}
% also when taking all orders of $\epsilon$ into account \cite{BBCK07}.
% The limit of $\xi_{1,2}$ for large $\xi_{2,1}$ is
% $(1-\theta)\omega^2$, which means that for initial conditions beyond
% order $1$ in the rescaled coordinate $\xi$ the map $\Pi_2$ is constant (as
% would be predicted by the quasi-static approximation).

When we combine the graph $\xi_{1,2}(\xi_{2,1})$ with the maps
$\Pi_1$, the projection $P_3$ and the map $\Pi_3$ we obtain an
implicit expression for the graph of the $P_\mathrm{DM}$ for
$\xi_{1,0}<0$, parametrized by the sliding time $t_2\in\R^+$:
\begin{align}
  \xi_{1,0}&=\ -\frac{\omega^2\theta^2}{2}\left[\frac{(1-\theta)\left[1-\exp(-t_2/\theta)\right]
      +t_2}{\theta+(1-\theta)\exp(-t_2/\theta)}\right]^2\label{eq:p:xi10t2}\\
  \xi_{1,4}&=\ \omega^2(1-\theta)\,
  \frac{\theta-\left[\theta+t_2\right]\exp(-t_2/\theta)}{\theta+(1-\theta)\exp(-t_2/\theta)}\mbox{.}
  \label{eq:p:xi14t2}
\end{align}
Its derivative at $\xi_{1,0}=0$ (that is, $t_2=0$) is
$(\theta-1)/\theta$. This is in line with the piecewise linearization
\eqref{eq:p:pdm}, and gives evidence that $\epsilon^2$ is really the
smallest present scale.

\section{Mechanics example --- autonomous dry-friction oscillator}
\label{sec:dfo1}
\begin{figure}[t]
  \centering
  \includegraphics[width=\textwidth]{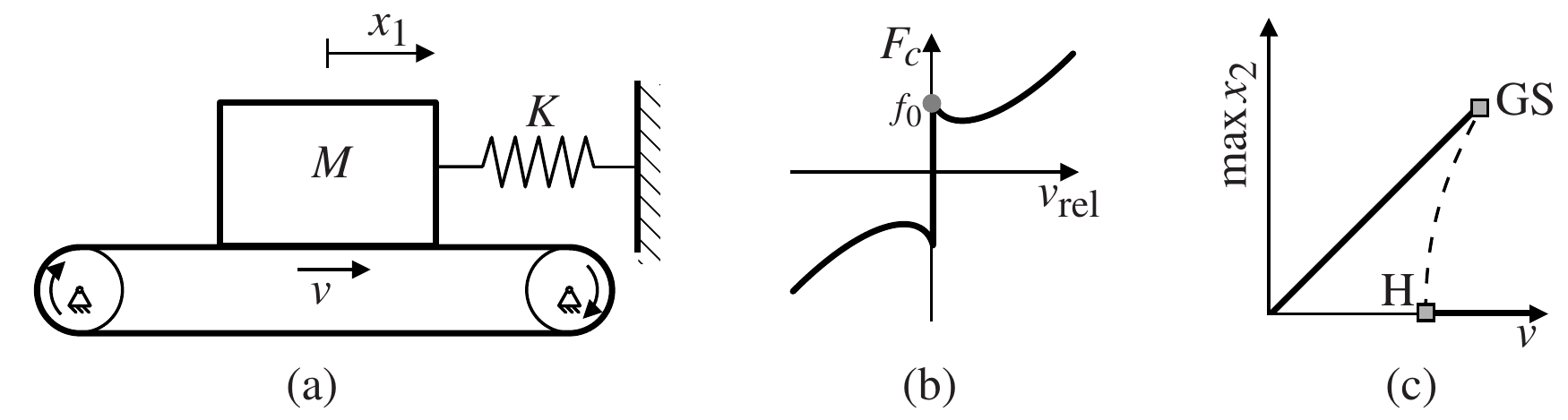}
  \caption{Dry friction oscillator: (a) setup, (b) friction force
    $F_c$ between belt and block depending on relative velocity
    $v_\mathrm{rel}=\dot x-v=x_2-v$, (c) bifurcation diagram as predicted by
    simple model \eqref{eq:dfo1} with a grazing-sliding bifurcation at
    $\mathrm{GS}$.}
  \label{fig:dfo}
\end{figure}
We demonstrate the practical consequences of Lemma~\ref{thm:gen} for a
simple mechanical system: the autonomous dry-friction oscillator as
shown in Figure~\ref{fig:dfo}(a), a block of mass $M$ attached to a
rigid wall via a spring of stiffness $K$ and slipping (or sticking) on
a belt moving with velocity $v$ toward the wall. This setup is a
caricature of the mechanics of, for example, violins or door squeal.
A simple model for the dry-friction oscillator is
\begin{equation}
  \label{eq:dfo1}
  \begin{split}
    \dot x_1&=x_2\\
    \dot x_2&=-\frac{K}{M}x_1-\frac{1}{M}F_c(x_2-v)
  \end{split}
\end{equation}
where $x_1$ is the position of the block, $x_2$ is its velocity and
$F_c$ is the friction force on the block exerted by the belt and
acting against the relative velocity. The switching between vector
fields is hidden in the form of the friction force $F_c$, which is the
sum of a smooth nonlinear function and a term $-f_0\sgn h$ where $h$ is
$h(x)=v-x_2$ for \eqref{eq:dfo1}. See Figure~\ref{fig:dfo}(b) for a
typical model of $F_c$. Figure~\ref{fig:dfo}(c) shows the bifurcation
diagram predicted by model~\eqref{eq:dfo1} (and qualitatively
confirmed in experiments in \cite{H00}), showing that the equilibrium
$x=(-F_c(-v)/K,0)$ loses its stability in a subcritical Hopf
bifurcation at $\mathrm{H}$ and that slip-stick oscillations exist and
are stable up to the belt velocity at point $\mathrm{GS}$. The
slip-stick oscillations correspond to periodic orbits with sliding
segment (physically corresponding to sticking: the block has zero
velocity relative to the belt for a while).
%Fc=-0.5;
%Fs=2;
%Fv=0.3;
%%friction definiton
%Ff=@(w)(Fc+(Fs-Fc)*exp(-abs(w))+Fv*abs(w));
\begin{figure}[t]
  \centering
   \subfigure[$\epsilon=0.1$, $v_0=3.32$]{\includegraphics%
    [width=0.48\columnwidth]{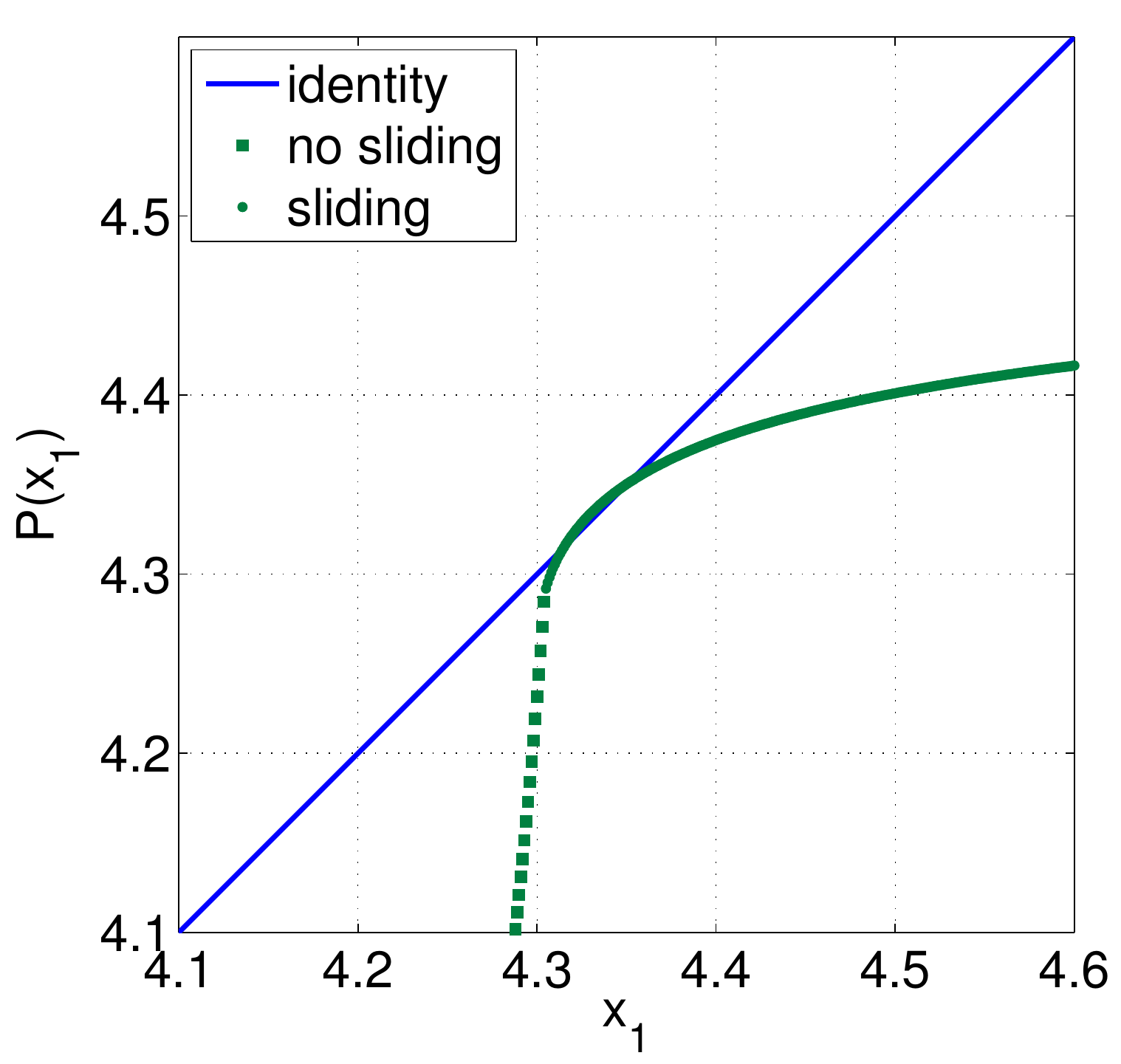}}
  \subfigure[$\epsilon=0.01$, $v_0=3.1$]{\includegraphics%
    [width=0.48\columnwidth]{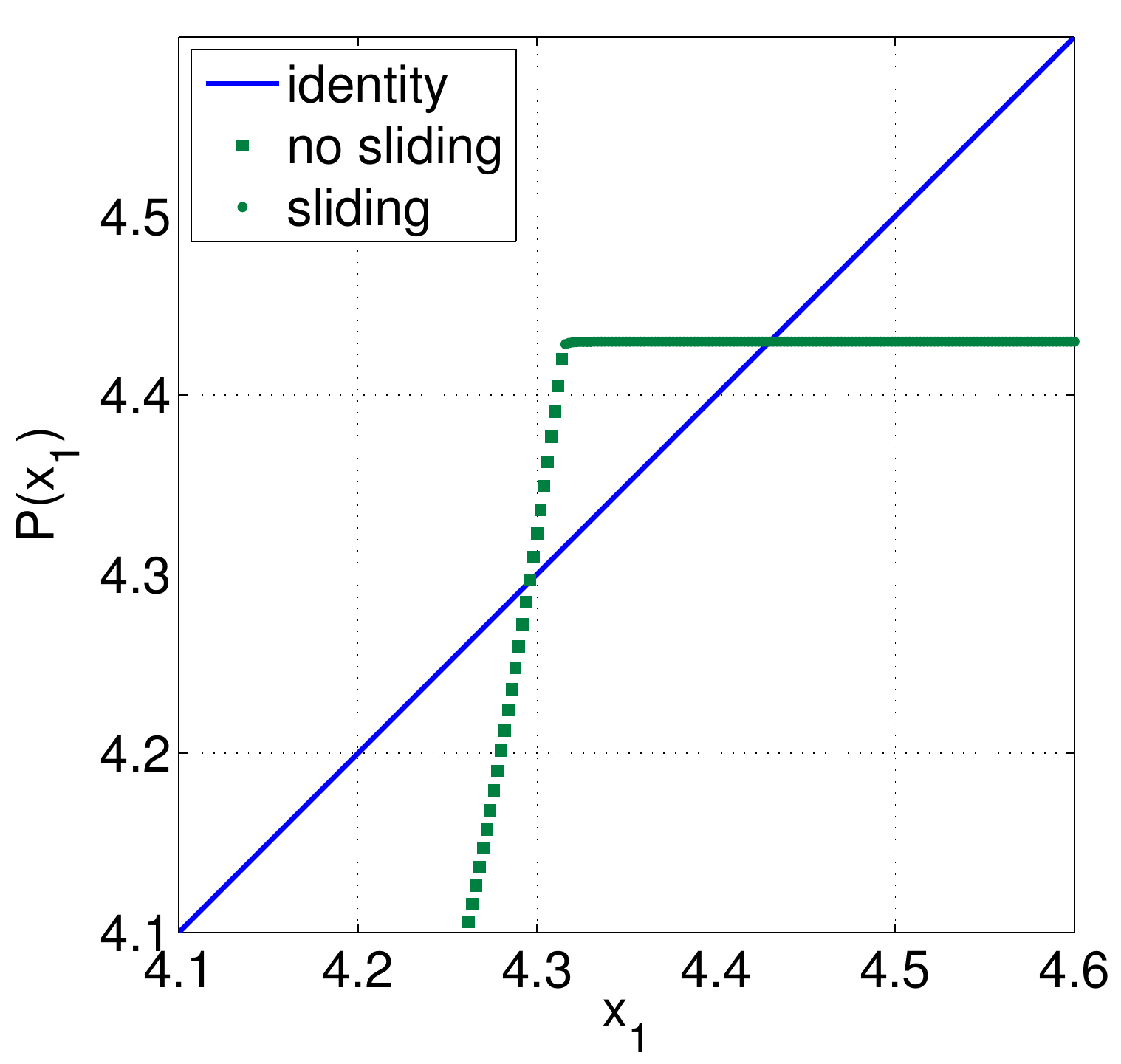}}
  \caption{Approximate return maps $P$ for the half-plane $x_2=0$,
    $x_1>0$ for dry friction oscillator \eqref{eq:dfo-sf}. Other
    (non-dimensional) parameters are $M=K=1$, $\gamma=2$,
    $F_c(\nu)=f_0\sgn\nu+f_v\nu+f_c\left(1-\ef^{-|\nu|}\right)\sgn\nu$
    where $f_0=-2$, $f_v=0.3$, and $f_c=2.5$.}
  \label{fig:dfo-retmap}
\end{figure}
At $\mathrm{GS}$ the
slip-stick oscillations lose their stability (and cease to exist) in a
grazing-sliding bifurcation as introduced in
Section~\ref{sec:gs-sgperturb} and discussed for the simple minimal
example in Section~\ref{sec:gs-min} (except that the orbits lying
entirely in $\mathcal{H}_+$ are unstable). Model~\eqref{eq:dfo1} can
be affected by singular perturbation in various ways. One possiblity
is that the velocity $v$ is not perfectly attained as a direct control
parameter but instead a controlled motor is aiming to achieve a
certain belt velocity $v_0$. This would imply that $v$ is determined
as a dependent variable of a fast subsystem, for example
\begin{equation}
  \label{eq:dfov}
  M_\mathrm{belt} \dot v=a[v_0-v]+ F_c(x_2-v)
\end{equation}
where $a$ is the proportional control gain and $M_\mathrm{belt}$ is
the effective mass of the belt. If we introduce the variable
$w=\left(M/M_\mathrm{belt}\right)x_2+v$ and the parameter
$\gamma=M/M_\mathrm{belt}$, and treat $M_\mathrm{belt}/a$ as the small
parameter $\epsilon$ then system~\eqref{eq:dfo1}, \eqref{eq:dfov} is
in the slow-fast form \eqref{eq:sgfilipslow}, \eqref{eq:sgfilipfast}
as described in the introduction. Namely we have
\begin{equation}
  \label{eq:dfo-sf}
  \begin{split}
    \dot x_1&=\ x_2\\
    \dot x_2&=\ -\frac{K}{M}x_1-\frac{1}{M}F_c((1+\gamma)x_2-w)\\
    \epsilon \dot w_{\phantom{2}} &=\ -w+v_0+\gamma x_2 -\epsilon
    \frac{K\gamma}{M} x_1\mbox{.}
\end{split}
\end{equation}
 The slow variable is $x=(x_1,x_2)$,
the fast variable is $w$, and the switching function $h$ is
$h(x,w)=w-(1+\gamma)x_2$.

The slow manifold $\mathcal{M}_0$ ($\epsilon=0$ in the fast subsystem
of \eqref{eq:dfo-sf}) is the graph of the function $w(x)=v_0+\gamma
x_2\}$. The first-order approximation of $\mathcal{M}_\pm$ is the
graph of $w_\pm(x)=v_0+\gamma x_2+\epsilon F_c(x_2-v_0)$. Thus, at the
switching surface we have that $w_\pm(x)=v_0+\gamma x\mp\epsilon
f_0/M$, which means that one effect of the imperfect control (the
non-zero $\epsilon$) is a slow-down of the belt velocity of order
$\epsilon$ in $\mathcal{H}_+$. The other, less obvious, effect is that
the non-zero $\epsilon$ can limit the minimal length of sliding
(sticking) segments that we can observe for the stable stick-slip
oscillations.  Figure~\ref{fig:dfo-retmap} shows the approximate
return map to the half-line $\mathcal{L}=\{(x,w): w=w_+(x), x_2=0,
x_1>0\}$ close to the grazing-sliding point $\mathrm{GS}$ shown in
Figure~\ref{fig:dfo}(c) (again, we have neglected the exponentially
small difference between $w$ and $w_+(x)$). The family of sliding
periodic orbits (the stick-slip oscillations) undergo a saddle-node
bifurcation and become unstable \emph{before} the sliding segment
shrinks to zero length. This is a consequence of the change of the
slope in the graph of the return map in Figure~\ref{fig:dfo-retmap}, a
change which is of order $1$ uniformly for $\epsilon\to0$. Apparently,
the grazing-sliding bifurcation in the dry-friction oscillator falls
into case 2 of Lemma~\ref{thm:gen}. Namely, we observe a grazing-sliding bifurcation and the periodic
orbit persists, but the stability of the periodic orbit may change
completely. The mechanism behind this change is identical to the case
$\theta>0$ of the minimal example in Lemma~\ref{thm:pdm}. We note that
the change of the right-side slopes of the return map in
Figure~\ref{fig:dfo-retmap} depends on $\gamma$ such that the
saddle-node bifurcation occurs only for sufficiently large $\gamma$
(that is, if the belt is not much heavier than the block).
\end{document}